\submitjournal{ApJ}

\shorttitle{Sample article}
\shortauthors{Zhang et al.}
\begin{document}

\title{Stellar chromospheric activity and age relation from open clusters in the LAMOST Survey}

\correspondingauthor{Jingkun Zhao}
\email{zjk@bao.ac.cn}

\author[0000-0002-0786-7307]{Jiajun Zhang}
\affil{Key Laboratory of Optical Astronomy, National Astronomical Observatories, Chinese Academy of Sciences, Beijing 100012, China. zjk@nao.cas.cn}
\affil{School of Astronomy and Space Science, University of Chinese Academy of Sciences, Beijing 100049, China}

\author{Jingkun Zhao}
\affiliation{Key Laboratory of Optical Astronomy, National Astronomical Observatories, Chinese Academy of Sciences, Beijing 100012, China. zjk@nao.cas.cn}

\author{Terry D. Oswalt}
\affiliation{Embry-Riddle Aeronautical University, 600 S. Clyde Morris Blvd., Daytona Beach FL, USA, 32114. oswaltt1@erau.edu}

\author{Xiangsong Fang}
\affiliation{Chinese Academy of Sciences South America Center for Astronomy, National Astronomical Observatories, CAS, Beijing 100101, China}
\affil{Instituto de Astronom\'{i}a, Universidad Cat\'{o}lica del Norte, Av. Angamos 0610, Antofagasta, Chile}

\author{Gang Zhao}
\affil{Key Laboratory of Optical Astronomy, National Astronomical Observatories, Chinese Academy of Sciences, Beijing 100012, China. zjk@nao.cas.cn}
\affil{School of Astronomy and Space Science, University of Chinese Academy of Sciences, Beijing 100049, China}

\author{Xilong Liang}
\affil{Key Laboratory of Optical Astronomy, National Astronomical Observatories, Chinese Academy of Sciences, Beijing 100012, China. zjk@nao.cas.cn}
\affil{School of Astronomy and Space Science, University of Chinese Academy of Sciences, Beijing 100049, China}

\author{Xianhao Ye}
\affil{Key Laboratory of Optical Astronomy, National Astronomical Observatories, Chinese Academy of Sciences, Beijing 100012, China. zjk@nao.cas.cn}
\affil{School of Astronomy and Space Science, University of Chinese Academy of Sciences, Beijing 100049, China}

\author{Jing Zhong}
\affiliation{Key Laboratory for Research in Galaxies and Cosmology, Shanghai Astronomical Observatory, Chinese Academy of Sciences, 80 Nandan Road, Shanghai 200030, China}



\begin{abstract}

We identify member stars of more than 90 open clusters in the LAMOST survey. With the method of \cite{FANG2018}, the chromospheric activity (CA) indices $\log R'\rm_{CaK}$ for 1091 member stars in 82 open clusters and $\log R'\rm_{H\alpha}$ for 1118 member stars in 83 open clusters are calculated. The relations between the average $\log R'\rm_{CaK}$, $\log R'\rm_{H\alpha}$ in each open cluster and its age are investigated in different $T\rm_{eff}$ and [Fe/H] ranges. We find that CA starts to decrease slowly from $\log t=6.70$ to $\log t=8.50$, and then decreases rapidly until $\log t=9.53$. The trend becomes clearer for cooler stars. The quadratic functions between $\log R'$ and $\log t$ with 4000K $<T\rm_{eff}<$ 5500K are constructed, which can be used to roughly estimate ages of field stars with accuracy about 40\% for $\log R'\rm_{CaK}$ and 60\% for $\log R'\rm_{H\alpha}$.

\end{abstract}



\section{Introduction} \label{sec:intro}
As a star ages, stellar rotation slows due to magnetic braking. In response, the magnetic field strength on stellar surface decreases. As the result, chromospheric heating drops. This paradigm is the current explanation for the observed decline in chromospheric activity (CA) with age \citep{Bab1961,Char2014}.

\cite{Sku1972} found that CaII HK lines emission decayed as the inverse square root of stellar age. Thus, CA is a potential age indicator. Several efforts have been undertaken to calibrate it. The quantity which is often used to indicate the strength of CA is $R\rm_{HK}'$. $R\rm_{HK}$ is the ratio of the flux in the core of CaII HK line to bolometric flux ($\sigma T\rm_{eff}^4$). $R\rm_{HK}$ is converted to $R\rm_{HK}'$ when photospheric contribution is removed. \cite{Sod1991} drew a linear relation between age ($\log t$) and CA ($\log R\rm_{HK}'$) using two open clusters and several visual binaries. They presumed that the relation was deterministic and not just statistical. Later, \cite{Roc1998} found a relationship between the observed difference between the stellar isochrone and chromospheric ages, and also the metallicity, as measured by the index [Fe/H] among late-type dwarfs. The chromospheric ages tended to be younger than the isochrone ages for metal-poor stars and the opposite occured for metal-rich stars. \cite{LAC1999} provided a CA vs. age relation for solar-type stars with $B-V>0.6$ using a piece-wise function.  Combining the cluster activity data with modern cluster age estimates, \cite{Mam2008} derived an improved activity-age calibration for F7-K2 dwarfs with 0.5 mag$ < B - V <$ 0.9 mag. \cite{PACE2004} used a sample of five open clusters and the sun to study the relation. They found an abrupt decay of CA occured between 0.6 Gyr and 1.5 Gyr, followed by a very slow decline. Later, \cite{PACE2013} found an L-shaped CA versus age diagram. They suggested the viability of this age indicator was limited to stars younger than about 1.5 Gyr. They detected no decay of CA after about 2 Gyr. However, \cite{Lor2016} took mass and [Fe/H] biases into account and established the viability of deriving usable chromospheric ages for solar-type stars up to at least ~6 Gyr. \cite{Lor2018} found evidence that, for the most homogenous set of old stars, CaII H and K activity indices seemed to continue decreasing after the solar age towards the lower main-sequence. Their results indicated that a significant part of the scatter observed in the age-activity relation of solar twins could be attributed to stellar cycle modulation effects.

The goal of this paper is to investigate the relations between CA and age in different $T\rm_{eff}$ and [Fe/H] ranges using the largest sample of open clusters in LAMOST \citep{cui2012,zhao2012}. Through these CA-age relations, we hope find a way to roughly estimate ages for main sequence stars in LAMOST, which are difficult to derive using the isochrone method. The paper is organized as follows. Section \ref{sec:data} describes data and sample. The measurements of CA indices $\log R'\rm_{CaK}$ and $\log R'\rm_{H\alpha}$ are presented in section \ref{sec:logr}. Our result and analysis are discussed in section \ref{sec:result}. Finally, our main conclusions are summarized in section \ref{sec:conclusion}.

\section{Data and sample} \label{sec:data}
\subsection{An overview of LAMOST}

The LAMOST telescope is a special reflecting Schmidt telescope \citep{cui2012,zhao2012,Luo2015}. Its primary mirror (Mb) is 6.67m$\times$6.05m and its correcting mirror (Ma) is 5.74m$\times$4.40m. It adopts an innovative active optics technique with 4,000 optical fibers placed on the focal surface. It can obtain spectra of 4,000 celestial objects simultaneously, which makes it the most efficient spectroscope in the world. In 2019 June, the LAMOST official website\footnote{\url{http://dr5.lamost.org/}} has released five data releases (DR5\_v3) to international astronomers. DR5\_v3 has 9,026,365 spectra for 8,183,160 stars, 153,863 galaxies, 52,453 quasars, and 637,889 unknown objects. These spectra cover the wavelength range of 3690-9100\AA\ with a resolution of 1800 at the 5500\AA. DR5\_v3 also provides stellar parameters such as effective temperature ($T\rm_{eff}$), metallicity ([Fe/H]), surface gravity ($\log g$) and radial velocity (RV) for millions of stars. The typical error for $T\rm_{eff}$, [Fe/H], $\log g$ and RV are 110K, 0.19dex, 0.11dex and 4.91 km/s, respectively \citep{Gao2015}. In this work, we measure CA indices $\log R'\rm_{CaK}$ and $\log R'\rm_{H\alpha}$ for the CaII K and H$\alpha$ lines. Our used spectra and stellar parameters including $T\rm_{eff}$, [Fe/H], $\log g$ and RV are all from DR5\_v3.

\subsection{Open clusters in LAMOST}

\cite{Gaudin2018} provided 401,448 member stars with membership probability of 1,229 open clusters in Gaia DR2. We select those member stars with membership probability $>$ 0.6. In addition, Melotte 25 (Hyades) is added from \cite{Roser2019}. The celestial coordinates of these member stars are used to cross match with the LAMOST general catalogue. Only dwarfs ($\log g>4.0$) with 4000K $<T\rm_{eff}<$ 7000K and signal-to-noise ratios (SNRs) satisfied some limits are selected. For a star with multiple spectra, only the spectrum with highest SNR is retained. For the CaII K line, 1,240 spectra of 89 open clusters with SNR g (signal-to-noise ratio in g band) $>$ 30 remain. For the H$\alpha$ line, 1,305 spectra of 93 open clusters with SNR r (signal-to-noise ratio in r band) $>$ 50 remain. Table \ref{clusters} lists these open clusters. The information about member stars can be found on online materials. The ages of these open clusters are from literatures as shown in Table \ref{clusters}. For most open clusters, their ages are from \cite{Kha2013}. However, the ages of eight open clusters are not found in literatures, which are not used to derive CA-age relations. Cluster ages are given as $\log t$, where $t$ is in units of yr. We calculate average [Fe/H] for each open cluster as shown in Table \ref{clusters}.

\startlongtable
\begin{deluxetable}{ccccccc}
\tablecaption{open clusters\label{clusters}}
\tablehead{
\colhead{name} & \colhead{J2000RA} & \colhead{J2000DEC} & \colhead{$\log t$} & \colhead{References}  & \colhead{$\rm [Fe/H]_{LA}$}  & \colhead{$\rm [Fe/H]\_std_{LA}$}
}
\startdata
  NGC\_2264 & 100.217 & 9.877 & 6.75 & 1 & 0.234 & \\
  Collinder\_69 & 83.792 & 9.813 & 6.76 & 1 & -0.014 & 0.1723\\
  IC\_348 & 56.132 & 32.159 & 6.78 & 1 & -0.024 & 0.1755\\
  NGC\_1333 & 52.297 & 31.31 & 6.8 & 1 & 0.196 & \\
  ASCC\_16 & 81.198 & 1.655 & 7.0 & 1 & -0.102 & 0.1223\\
  Stock\_8 & 81.956 & 34.452 & 7.05 & 1 & -1.189 & 1.0445\\
  ASCC\_21 & 82.179 & 3.527 & 7.11 & 1 & 0.053 & 0.0515\\
  Collinder\_359 & 270.598 & 3.26 & 7.45 & 1 & 0.034 & 0.1754\\
  ASCC\_19 & 81.982 & -1.987 & 7.5 & 1 & -0.123 & 0.088\\
  Stephenson\_1 & 283.568 & 36.899 & 7.52 & 1 & 0.2 & 0.0565\\
  NGC\_1960 & 84.084 & 34.135 & 7.565 & 1 & -0.246 & 0.0853\\
  Alessi\_20 & 2.593 & 58.742 & 7.575 & 1 & 0.057 & 0.1974\\
  Dolidze\_16 & 78.623 & 32.707 & 7.6 & 1 & -0.012 & 0.036\\
  IC\_4665 & 266.554 & 5.615 & 7.63 & 1 & 0.09 & 0.0831\\
  Melotte\_20 & 51.617 & 48.975 & 7.7 & 1 & 0.041 & 0.1049\\
  NGC\_2232 & 96.888 & -4.749 & 7.7 & 1 & 0.019 & 0.1087\\
  ASCC\_13 & 78.255 & 44.417 & 7.71 & 1 & -0.086 & \\
  ASCC\_114 & 324.99 & 53.997 & 7.75 & 1 & -0.127 & \\
  ASCC\_6 & 26.846 & 57.722 & 7.8 & 1 & -0.039 & \\
  FSR\_0904 & 91.774 & 19.021 & 7.8 & 1 & -0.178 & 0.0585\\
  Alessi\_19 & 274.741 & 12.311 & 7.9 & 1 & 0.073 & 0.0512\\
  ASCC\_105 & 295.548 & 27.366 & 7.91 & 1 & 0.059 & 0.0405\\
  Trumpler\_2 & 39.232 & 55.905 & 7.925 & 1 & -0.027 & 0.0365\\
  ASCC\_113 & 317.933 & 38.638 & 7.93 & 1 & -0.011 & \\
  NGC\_7063 & 321.122 & 36.507 & 7.955 & 1 & -0.093 & 0.0064\\
  NGC\_7243 & 333.788 & 49.83 & 7.965 & 1 & -0.068 & 5.0E-4\\
  ASCC\_29 & 103.571 & -1.67 & 8.06 & 1 & -0.234 & 0.008\\
  NGC\_7086 & 322.624 & 51.593 & 8.065 & 1 & 0.004 & \\
  Alessi\_37 & 341.961 & 46.342 & 8.125 & 2 & 0.057 & 0.0213\\
  Melotte\_22 & 56.601 & 24.114 & 8.15 & 1 & 0.006 & 0.1241\\
  Alessi\_Teutsch\_11 & 304.127 & 52.051 & 8.179 & 2 & 0.029 & \\
  NGC\_2186 & 93.031 & 5.453 & 8.2 & 1 & -0.233 & \\
  NGC\_2168 & 92.272 & 24.336 & 8.255 & 1 & -0.062 & 0.0759\\
  Gulliver\_20 & 273.736 & 11.082 & 8.289 & 2 & -0.097 & 0.013\\
  FSR\_0905 & 98.442 & 22.312 & 8.3 & 1 & -0.154 & \\
  NGC\_1647 & 71.481 & 19.079 & 8.3 & 1 & 0.025 & 0.0718\\
  ASCC\_11 & 53.056 & 44.856 & 8.345 & 1 & -0.081 & 0.037\\
  NGC\_1912 & 82.167 & 35.824 & 8.35 & 1 & -0.162 & 0.0342\\
  NGC\_2301 & 102.943 & 0.465 & 8.35 & 1 & -0.02 & 0.0675\\
  NGC\_744 & 29.652 & 55.473 & 8.375 & 1 & -0.213 & \\
  NGC\_1039 & 40.531 & 42.722 & 8.383 & 1 & -0.027 & 0.1309\\
  Stock\_10 & 84.808 & 37.85 & 8.42 & 1 & -0.09 & 0.0736\\
  ASCC\_108 & 298.306 & 39.349 & 8.425 & 1 & -0.046 & 0.0589\\
  Stock\_2 & 33.856 & 59.522 & 8.44 & 1 & -0.056 & 0.0694\\
  Czernik\_23 & 87.525 & 28.898 & 8.48 & 1 & 0.157 & \\
  ASCC\_23 & 95.047 & 46.71 & 8.485 & 1 & -0.032 & 0.074\\
  FSR\_0985 & 92.953 & 7.02 & 8.5 & 1 & 0.056 & \\
  Stock\_1 & 294.146 & 25.163 & 8.54 & 1 & 0.048 & \\
  NGC\_1528 & 63.878 & 51.218 & 8.55 & 1 & -0.145 & 0.0415\\
  NGC\_2099 & 88.074 & 32.545 & 8.55 & 1 & -0.017 & 0.0549\\
  Ferrero\_11 & 93.646 & 0.637 & 8.554 & 2 & -0.103 & 0.044\\
  NGC\_7092 & 322.889 & 48.247 & 8.569 & 1 & -0.29 & \\
  NGC\_1342 & 52.894 & 37.38 & 8.6 & 1 & -0.155 & 0.0809\\
  NGC\_1907 & 82.033 & 35.33 & 8.6 & 1 & -0.18 & \\
  NGC\_2184 & 91.69 & -2.0 & 8.6 & 1 & -0.074 & 0.082\\
  NGC\_1750 & 75.926 & 23.695 & 8.617 & 2 & -0.017 & 0.0846\\
  ASCC\_12 & 72.4 & 41.744 & 8.63 & 1 & -0.085 & \\
  NGC\_6866 & 300.983 & 44.158 & 8.64 & 1 & 0.019 & 0.067\\
  NGC\_1582 & 67.985 & 43.718 & 8.665 & 1 & -0.072 & 0.0742\\
  Roslund\_6 & 307.185 & 39.798 & 8.67 & 1 & 0.024 & 0.0797\\
  NGC\_1662 & 72.198 & 10.882 & 8.695 & 1 & -0.111 & 0.0865\\
  Alessi\_2 & 71.602 & 55.199 & 8.698 & 1 & -0.01 & 0.0618\\
  ASCC\_41 & 116.674 & 0.137 & 8.7 & 1 & -0.093 & 0.0774\\
  Collinder\_350 & 267.018 & 1.525 & 8.71 & 1 & -0.034 & 0.0923\\
  ASCC\_10 & 51.87 & 34.981 & 8.717 & 1 & -0.02 & 0.049\\
  NGC\_2548 & 123.412 & -5.726 & 8.72 & 1 & -0.026 & 0.0624\\
  NGC\_1758 & 76.175 & 23.813 & 8.741 & 2 & -0.013 & 0.0523\\
  NGC\_1664 & 72.763 & 43.676 & 8.75 & 1 & -0.082 & 0.0602\\
  NGC\_1708 & 75.871 & 52.851 & 8.755 & 1 & -0.065 & 0.0236\\
  NGC\_6633 & 276.845 & 6.615 & 8.76 & 1 & -0.098 & 0.0373\\
  NGC\_2281 & 102.091 & 41.06 & 8.785 & 1 & -0.033 & 0.1\\
  IC\_4756 & 279.649 & 5.435 & 8.79 & 1 & -0.087 & 0.0803\\
  NGC\_2194 & 93.44 & 12.813 & 8.8 & 1 & -0.008 & \\
  NGC\_1545 & 65.202 & 50.221 & 8.81 & 1 & -0.001 & \\
  Dolidze\_8 & 306.129 & 42.3 & 8.855 & 1 & 0.025 & \\
  Melotte\_25 & 66.725 & 15.87 & 8.87 & 3 & -0.003 & 0.13\\
  NGC\_1817 & 78.139 & 16.696 & 8.9 & 1 & -0.205 & 0.0936\\
  NGC\_2355 & 109.247 & 13.772 & 8.9 & 1 & -0.248 & 0.0538\\
  NGC\_2632 & 130.054 & 19.621 & 8.92 & 1 & 0.187 & 0.1053\\
  King\_6 & 51.982 & 56.444 & 8.975 & 1 & -0.055 & \\
  NGC\_6811 & 294.34 & 46.378 & 9.0 & 4 & -0.08 & 0.0768\\
  NGC\_1245 & 48.691 & 47.235 & 9.025 & 1 & -0.186 & \\
  NGC\_752 & 29.223 & 37.794 & 9.13 & 1 & -0.067 & 0.0825\\
  Koposov\_63 & 92.499 & 24.567 & 9.22 & 1 & -0.016 & \\
  NGC\_7789 & -0.666 & 56.726 & 9.265 & 1 & -0.145 & 0.1199\\
  NGC\_2112 & 88.452 & 0.403 & 9.315 & 1 & -0.138 & 0.0311\\
  NGC\_2420 & 114.602 & 21.575 & 9.365 & 1 & -0.278 & 0.0462\\
  NGC\_2682 & 132.846 & 11.814 & 9.535 & 1 & -0.003 & 0.0566\\
  Gulliver\_22 & 84.848 & 26.368 &  &  & -0.225 & \\
  Gulliver\_25 & 52.011 & 45.152 &  &  & -0.009 & \\
  Gulliver\_6 & 83.278 & -1.652 &  &  & -0.049 & 0.1697\\
  Gulliver\_60 & 303.436 & 29.672 &  &  & -0.12 & 0.002\\
  Gulliver\_8 & 80.56 & 33.792 &  &  & -0.266 & 0.0745\\
  RSG\_1 & 75.508 & 37.475 &  &  & -0.014 & 0.0834\\
  RSG\_5 & 303.482 & 45.574 &  &  & 0.098 & 0.093\\
  RSG\_7 & 344.19 & 59.363 &  &  & 0.014 & 0.012\\
\enddata
\tablecomments{The first column is name of open clusters. The second and third columns are mean right ascension and decination (J2000) of member stars and they are in units of ($^{\circ}$). Mean right ascension and decination of all clusters but Melotte 25 are from \cite{Gaudin2018}, while the coordinates of Melotte 25 are from \cite{Dias2014}. The fourth column is ages of clusters which are represented by $\log t$, where $t$ is in units of yr. The fifth column is references from which the ages are cited: 1-\cite{Kha2013}, 2-\cite{Bossini2019}, 3-\cite{Gos2018}, 4-\cite{Sandquist2016}. However, the ages of eight open clusters are not found in literatures. The sixth and seventh columns are the mean value and standard deviation of [Fe/H] of each open cluster. Some clusters have no standard deviation, which means they are represented by only one member star. Note that Stock 8 has $\rm [Fe/H]_{LA}=-1.189$. The cluster has only two member stars, of which one has $\rm [Fe/H]=-2.23$. This star shouldn't be a member star of the cluster. We don't calculate its $\log R'$ values.}
\end{deluxetable}

\section{Determination of excess fractional luminosity} \label{sec:logr}

Fang's method \citep{FANG2018} is used to calculate the excess fractional luminosities $\log R\rm'_{CaK}$ for the CaII K line and $\log R\rm'_{H\alpha}$ for the H$\alpha$ line. We use only the CaII K line because the CaII H line might be polluted by a hydrogen line. First, equivalent width ($\rm EW$) for the CaII K line and the H$\alpha$ line are measured. Then, excess equivalent width ($\rm EW'$) are obtained. As a last step, we calculate $\chi$, the ratio of the surface continuum flux near the line to the stellar surface bolometric flux from model spectra. We then compute the excess fractional luminosity $\log R'$.

\subsection{Measurement of equivalent width} \label{subsec:EW}
\begin{equation}\label{sec:EW}
\mathrm{EW}=\int \frac{f(\lambda)-f(\lambda_{c})}{f(\lambda_{c})}d\lambda
\end{equation}

As an example, the $\rm EW$ of CaII K line is measured by using Equation \ref{sec:EW}. Here, $f(\lambda_{c})$ denotes the pseudo-continuum flux. To measure the $\rm EW$ of the CaII K line, we integrate the line flux from 3930.2\AA\ to 3937.2\AA. The pseudo-continuum flux $f(\lambda_{c})$ is estimated by interpolating the flux between 3905.0-3920.0\AA\ and 3993.5-4008.5\AA. These values for CaII K and H$\alpha$ can be found in Table \ref{tab:wavelength regimes of EW}.

\begin{figure}[ht!]
\centering
\includegraphics[scale=0.7]{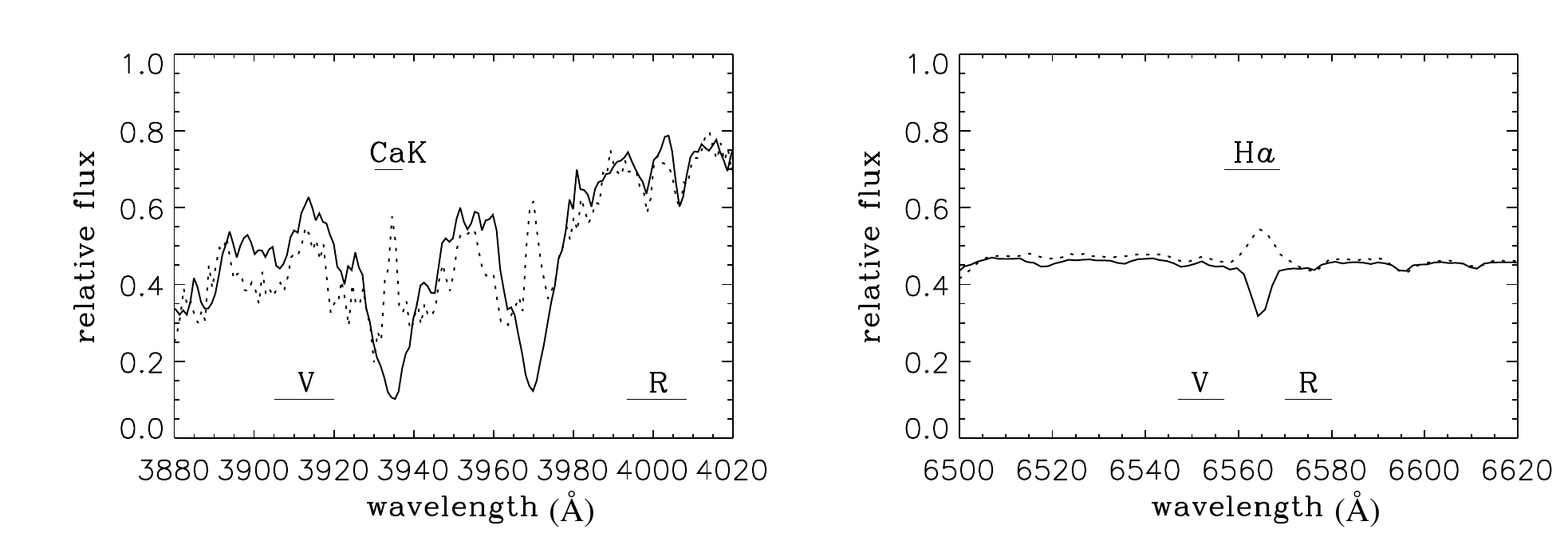}
\caption{Spectrum of an active star (dotted line) and an inactive star (solid line). The solid lines under the letters indicate the wavelength regimes used to measure $\rm EW$. \label{fig:active and inactive}}
\end{figure}

\begin{figure}
\gridline{\fig{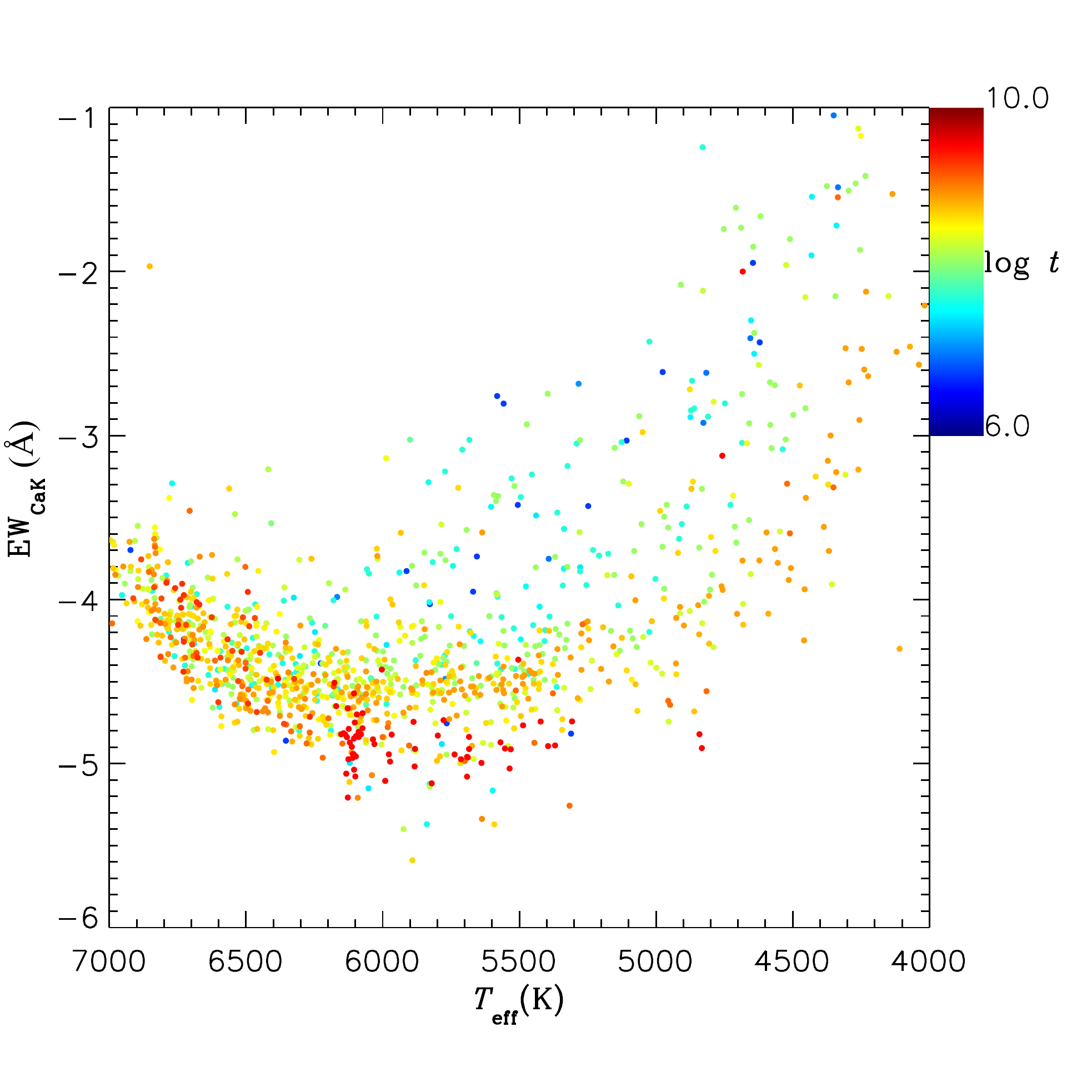}{0.52\textwidth}{(a)}
          \fig{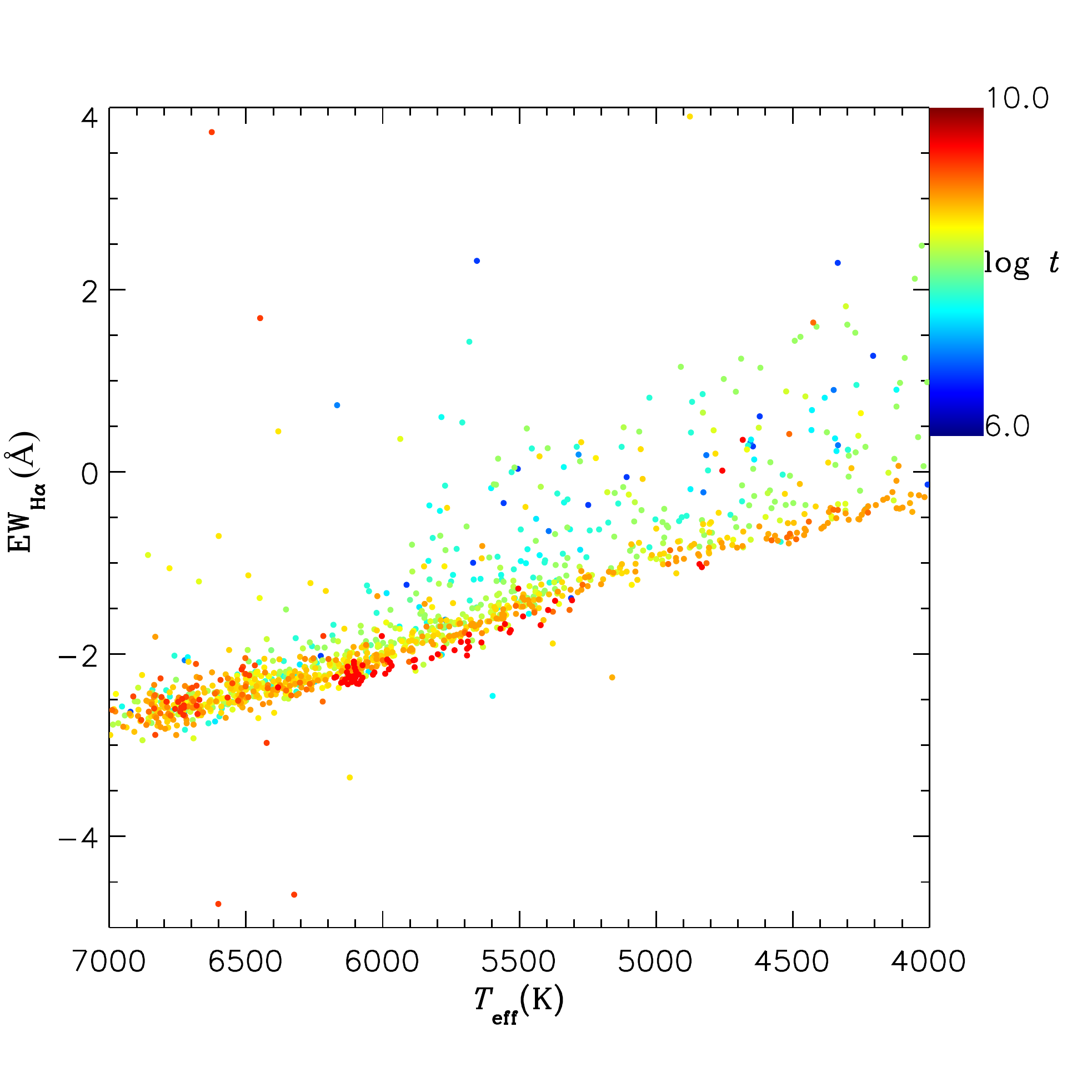}{0.52\textwidth}{(b)}
          }
\caption{(a): $\rm EW_{CaK}$ vs. $T\rm_{eff}$. (b): $\rm EW_{H\alpha}$ vs. $T\rm_{eff}$. Color is used to represent for the ages of open clusters.   \label{fig: EW_Teff}}
\end{figure}

Figure \ref{fig:active and inactive} shows the spectra of an active star (dotted line) and an inactive star (solid line). V and R represent wavelengths of violet and red pseudo-continua, respectively. The labels 'CaK' and 'H$\alpha$' illustrate the wavelength regimes of CaII K and H$\alpha$ lines, respectively. $\rm EW$ are measured from radial velocity corrected spectra. Figure \ref{fig: EW_Teff} plots $\rm EW_{CaK}$ vs. $T\rm_{eff}$ and $\rm EW_{H\alpha}$ vs. $T\rm_{eff}$. Color is used to represent for the ages of open clusters.

We use simple Monte Carlo simulation to obtain the error of $\rm EW$. Detail information can be found in Appendix \ref{Measurement error of logR}. For the CaII K line, the average error of $\rm EW_{CaK}$ is about 0.1\AA\ when $T\rm_{eff}>5500$K. The average error of $\rm EW_{CaK}$ is about 0.2\AA\ at $T\rm_{eff}=4500$K. For the H$\alpha$ line, the average error of $\rm EW_{H\alpha}$ is about 0.04\AA.

From Figure \ref{fig: EW_Teff} (a), it is clear that as $T\rm_{eff}$ decreases, $\rm EW_{CaK}$ first decreases and then increases. Figure \ref{fig: EW_Teff} (b) shows that as $T\rm_{eff}$ decreases $\rm EW_{H\alpha}$ increases. As for the two panels, at high $T\rm_{eff}$ range ($T\rm_{eff}>6500$K), member stars of different age populations mix. As $T\rm_{eff}$ decreases, young member stars tend to have larger $\rm EW$ than old member stars, which conforms to our expectation. The scatter of $\rm EW$ increases as $T\rm_{eff}$ decreases. Most member stars in our sample have $T\rm_{eff}>5500$K.

\startlongtable
\begin{deluxetable}{ccc}
\tablecaption{Equivalent width measurements of CaII K and H$\alpha$ \label{tab:wavelength regimes of EW}}
\tablehead{
\colhead{Line} & \colhead{Line bandpass(\AA)} & \colhead{Pseudo-continua(\AA)}
}
\startdata
CaII K & 3930.2-3937.2 & 3905.0-3920.0, 3993.5-4008.5 \\
H$\alpha$ & 6557.0-6569.0 & 6547.0-6557.0, 6570.0-6580.0 \\
\enddata
\end{deluxetable}

\subsection{Excess equivalent width}

\begin{figure}
\gridline{\fig{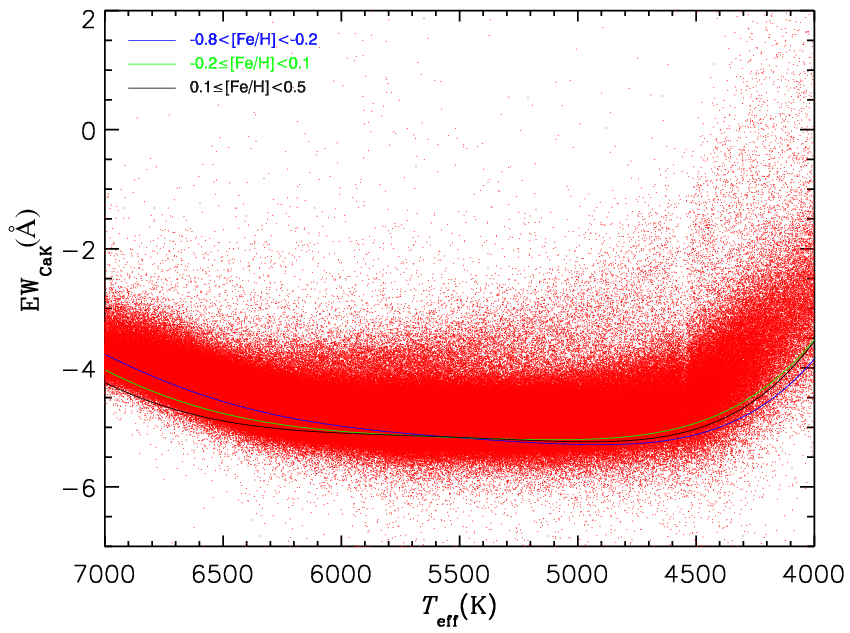}{0.52\textwidth}{(a)}
          \fig{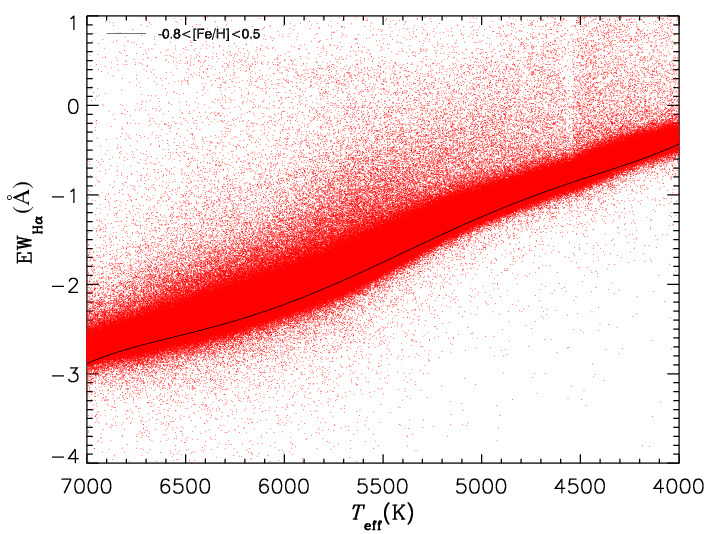}{0.52\textwidth}{(b)}
          }
\caption{(a): $\rm EW_{CaK}$ vs. $T\rm_{eff}$ for stars in the LAMOST AFGK type star catalog. (b): $\rm EW_{H\alpha}$ vs. $T\rm_{eff}$ for stars in the same catalog. The scarcity of stars at about 4570K may be caused by a defect in LAMOST pipeline at this $T\rm_{eff}$. The solid lines in the two panels are our basal lines. The left panel shows three basal lines corresponding to three different [Fe/H] ranges. \label{fig:inactivesample}}
\end{figure}

To measure excess equivalent width $\rm EW'_{CaK}$ and $\rm EW'_{H\alpha}$, basal lines are needed. The LAMOST official website provides the AFGK type star catalog. In this catalog, stars which satisfy 4000K $<T\rm_{eff}<$ 7000K, $\log g>4.0$ and SNR g $>$ 30 (SNR r $>$ 50), the same limitations as member stars, are selected. Futher, only stars with -0.8 $<$ [Fe/H] $<$ 0.5 are selected because at poor [Fe/H] range ([Fe/H] $\leq$ -0.8) there is a negative correlation between $\rm EW_{CaK}$ and [Fe/H]. We calculate their $\rm EW_{CaK}$ and $\rm EW_{H\alpha}$ by the same method described in section \ref{subsec:EW}. Those stars with $\rm EW > 10\AA$ or $\rm EW < -10\AA$ are excluded.  Figure \ref{fig:inactivesample} shows $\rm EW_{CaK}$ vs. $T\rm_{eff}$ and $\rm EW_{H\alpha}$ vs. $T\rm_{eff}$. This plot includes 1,563,898 stars for $\rm EW_{CaK}$ and 1,581,197 stars for $\rm EW_{H\alpha}$. Few stars are located at about 4570K. This may be caused by a defect in the LAMOST pipeline at this $T\rm_{eff}$.

Because [Fe/H] has more effects on $\rm EW_{CaK}$ than $\rm EW_{H\alpha}$, we classify stars into three classes according to [Fe/H]: $-0.8<$ [Fe/H] $<-0.2$, $-0.2\leq$ [Fe/H] $<0.1$ and $0.1\leq$ [Fe/H] $<0.5$ before determining the basal lines for $\rm EW_{CaK}$. For each class, we rebin the stars on $T\rm_{eff}$  with a bin width of 50 K. In each bin, 10\% quantile is calculated in $\rm EW_{CaK}$. Then five order polynomial is fitted to these quantiles as shown in the left panel of Figure \ref{fig:inactivesample}. The solid lines are fitting curves. From the left panel we can see that the three basal lines corresponding to three different [Fe/H] ranges have some differences. When $T\rm_{eff}>6000$K, the basal line of poor [Fe/H] range is above on that of rich [Fe/H] range. Generally speaking, poor [Fe/H] stars have larger $\rm EW_{CaK}$ than rich [Fe/H] stars because for poor [Fe/H] stars metallic lines are relatively shallow. Note that our $\rm EW$ is negative for an absorption line. For H$\alpha$, we directly rebin the stars on $T\rm_{eff}$ with a bin width of 50 K without [Fe/H] classification and the basal line is obtained by the same method as above. The right panel of Figure \ref{fig:inactivesample} shows the basal line for $\rm EW_{H\alpha}$. Then for member stars, $\rm EW'_{CaK}$ and $\rm EW'_{H\alpha}$ are obtained by using Equation \ref{sec:EWpie}. Stars whose $\rm EW'>0$ are retained in our study.

\begin{equation}\label{sec:EWpie}
\rm EW'=EW-EW_{basal}
\end{equation}

\subsection{Excess fractional luminosity}

After obtaining $\rm EW'$, Equations \ref{sec:chi} and \ref{sec:Rpie} are used to calculate the excess fractional luminosity $R'$. In Equation \ref{sec:chi}, ATLAS9 model atmospheres\footnote{\url{wwwuser.oats.inaf.it/castelli/grids.html}} are used to calculate $\chi$ of grid points. $F(\lambda)$ is flux at the stellar surface. We choose $\lambda=3950.5$\AA\ for CaII K and $\lambda=6560.0$\AA\ for H$\alpha$. $\sigma$ is Stefan-Boltzmann constant. Turbulent velocity (vturb = 2.0km/s) and mixing length parameter (1/H = 1.25) are adopted for the model spectra. We calculate $\chi$ in terms of $T\rm_{eff}$, [Fe/H] and $\log g$. $T\rm_{eff}$ ranges from 4000K to 7000K in steps of 250K. The values of $\log g$ range from 4.0 to 5.0 in steps of 0.5. [Fe/H] ranges from -2.5 to 0.5 in steps of 0.5 plus a value [Fe/H] = 0.2. Three dimensional linear interpolation is used to obtain the corresponding $\chi$ for each star. Finally, excess fractional luminosity $R'$ is obtained by using Equation \ref{sec:Rpie}.

\begin{equation}\label{sec:chi}
\chi=\frac{F(\lambda)}{\sigma T\rm_{eff}^4}
\end{equation}

\begin{equation}\label{sec:Rpie}
R\rm'=EW'\times \chi
\end{equation}

\section{Results and analysis} \label{sec:result}
\subsection{the distribution of $\log R'$}
\begin{figure}
\gridline{\fig{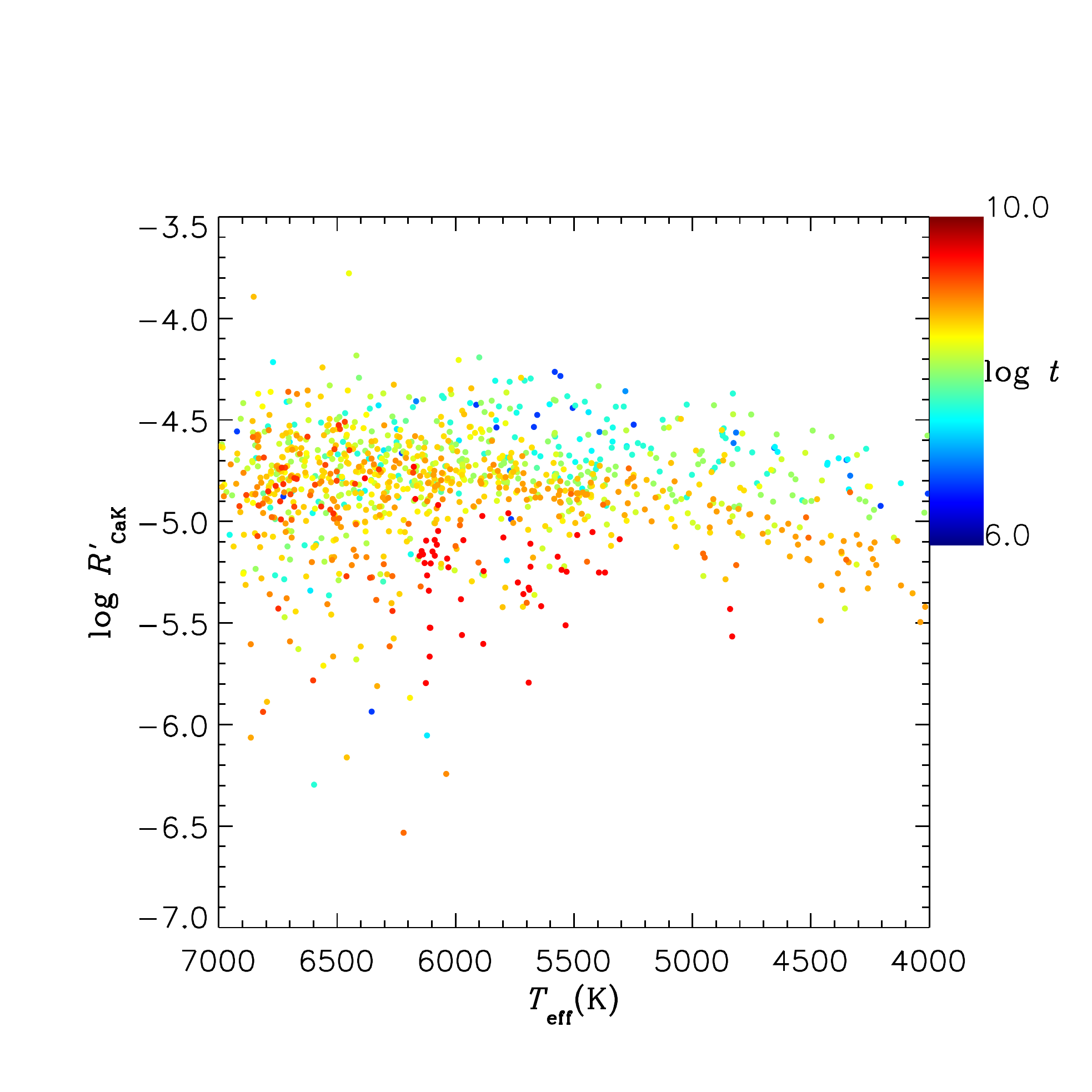}{0.52\textwidth}{(a)}
          \fig{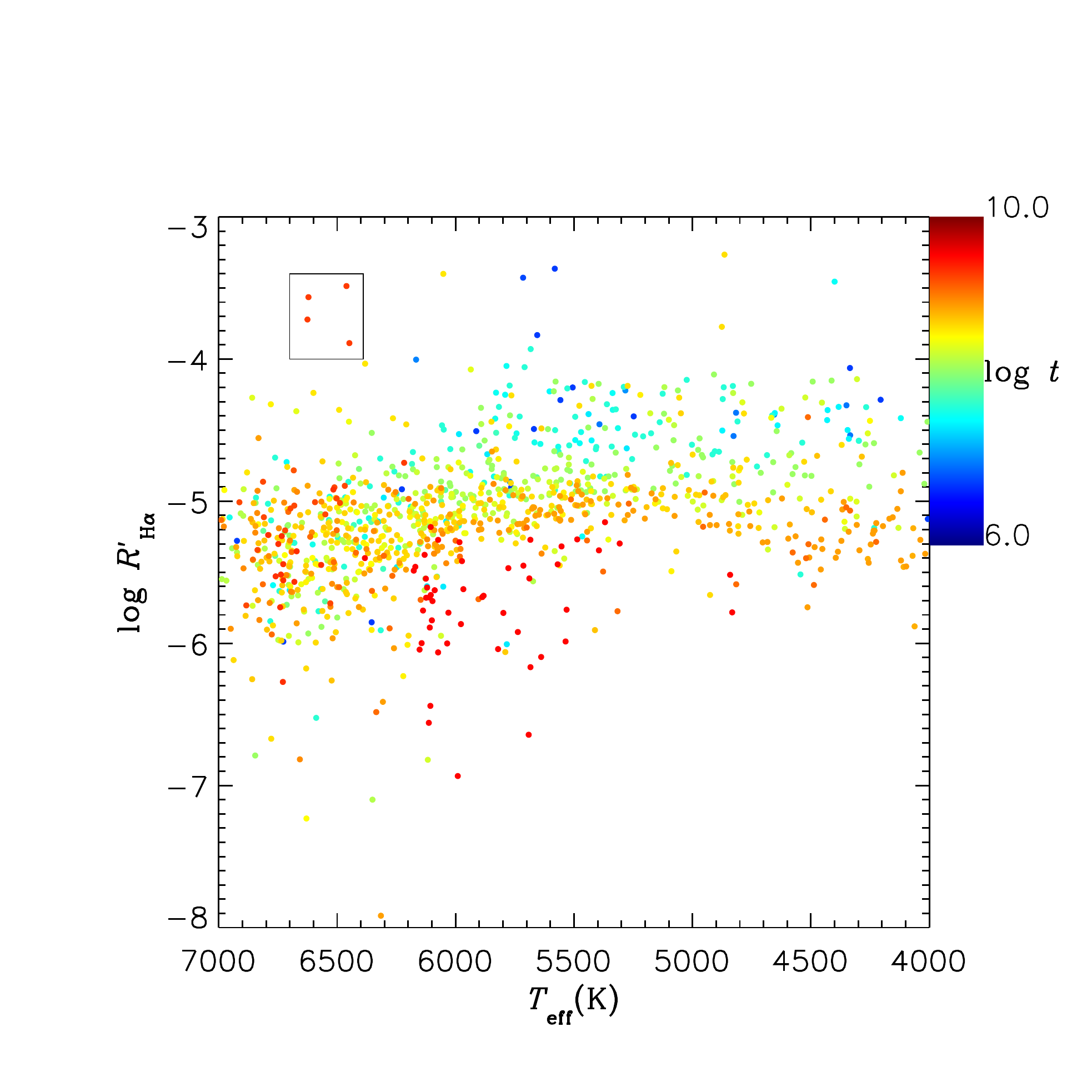}{0.52\textwidth}{(b)}
          }
\gridline{\fig{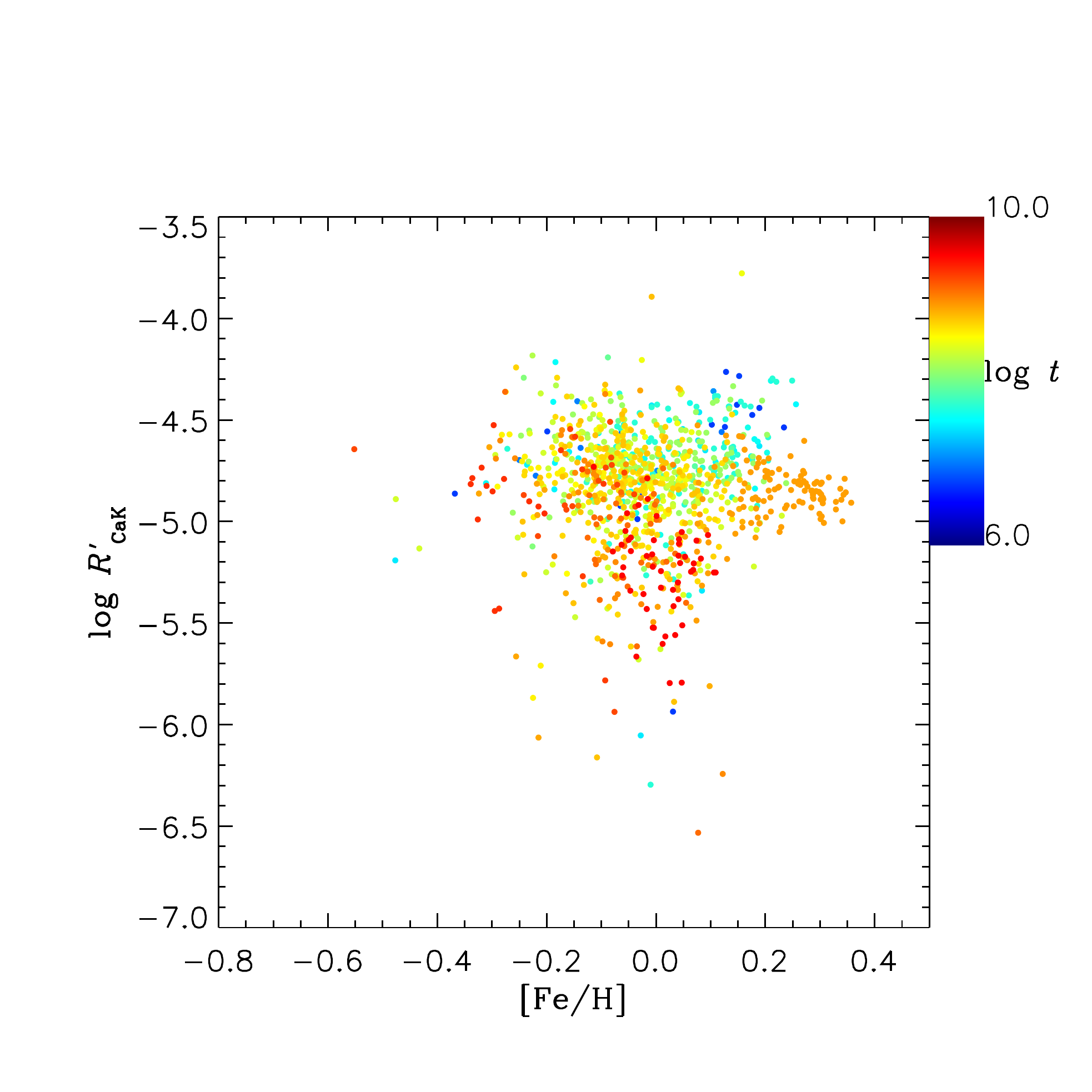}{0.52\textwidth}{(c)}
          \fig{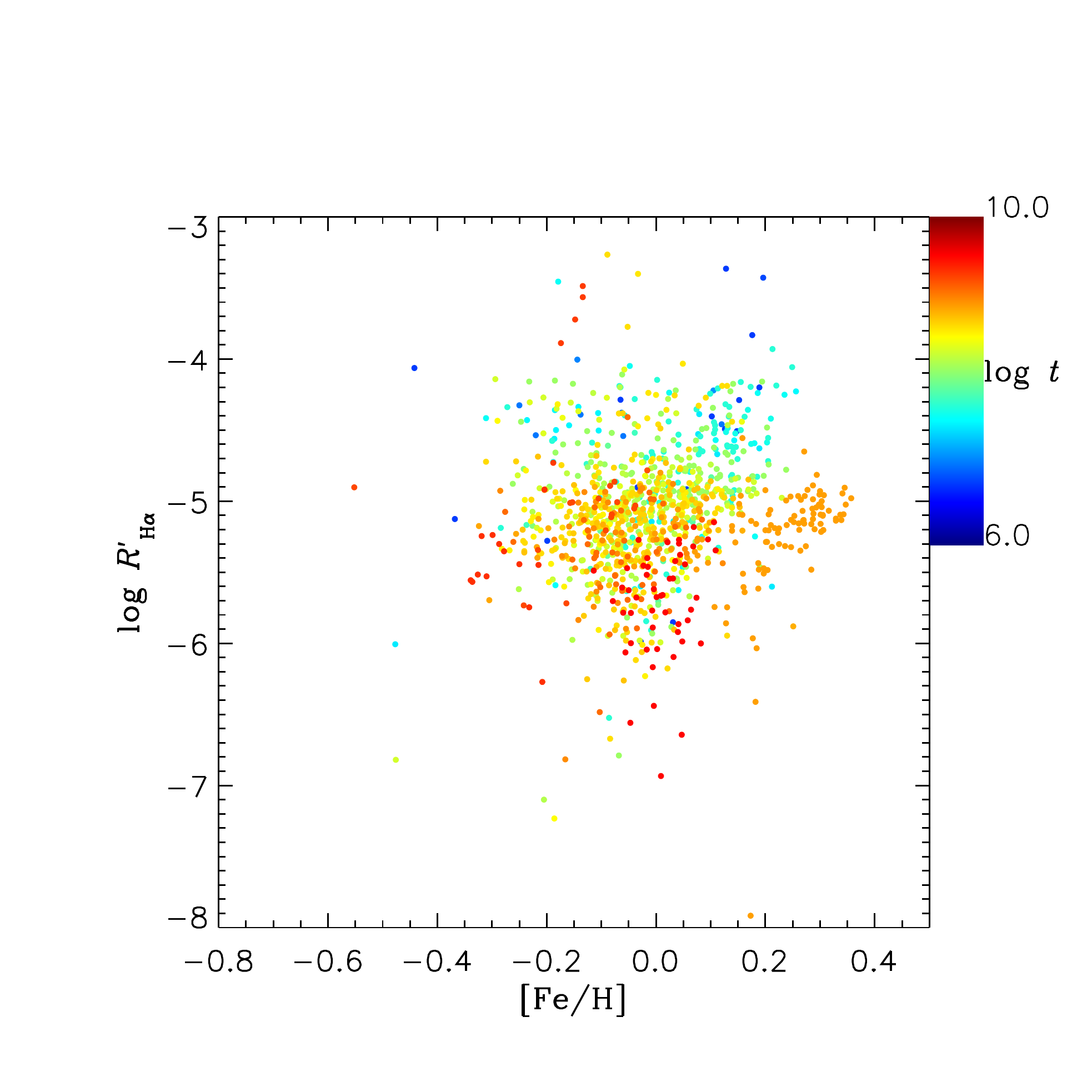}{0.52\textwidth}{(d)}
          }
\caption{(a): $\log R\rm'_{CaK}$ vs. $T\rm_{eff}$. (b): $\log R\rm'_{H\alpha}$ vs. $T\rm_{eff}$. (c): $\log R\rm'_{CaK}$ vs. [Fe/H]. (d): $\log R\rm'_{H\alpha}$ vs. [Fe/H]. The color has the same meaning as Figure \ref{fig: EW_Teff}. This plot includes 1091 member stars of 82 open clusters for CaII K and 1118 member stars of 83 open clusters for H$\alpha$, respectively.  \label{fig: logr_Teff}}
\end{figure}

With the above procedures, $\log R'$ value for each member star is obtained. We perform a cross-match between our sample and the Simbad database\footnote{\url{http://simbad.u-strasbg.fr/simbad/}}, and exclude those stars labeled as 'Flare*', 'pMS*', 'RSCVn', 'SB*', 'EB*WUMa', 'EB*', 'EB*Algol' and 'EB*betLyr' by Simbad. Flare stars and binaries can affect CA level \citep{Curtis2017,FANG2018}. In Appendix \ref{binaries and interstellar medium}, we simply discuss the impact of binaries. The result is shown in Figure \ref{fig: logr_Teff}. There are 1091 member stars in 82 open clusters with $\log R\rm'_{CaK}$ values and 1118 member stars in 83 open clusters with $\log R\rm'_{H\alpha}$ values. Some open clusters are represented by only one or two stars. The Melotte 22 and NGC 2632 have over 100 member stars. From Figure \ref{fig: logr_Teff}(a) and (b), young member stars have similar $\log R'$ values as old member stars when $T\rm_{eff}>6500$K. As $T\rm_{eff}$ decreases, young member stars tend to have larger $\log R'$ than old member stars, which conforms to our expectation. This phenomena is alike as Figure \ref{fig: EW_Teff}.

We use simple Monte Carlo simulation to obtain the error of $\log R'$. Detail information can be found in Appendix \ref{Measurement error of logR}. For the CaII K line, $\rm \sigma(\log R'_{CaK})$ has a large scatter when $T\rm_{eff}>6000$K (see Figure \ref{fig: EWerr_Teff}). $\rm \sigma(\log R'_{CaK})$ is about 0.05dex and 0.15dex at 5500K and 4500K. For the H$\alpha$ line, the distribution of $\rm \sigma(\log R'_{H\alpha})$ has a large scatter from 0.0dex to 0.5dex at all $T\rm_{eff}$ range (see Figure \ref{fig: EWerr_Teff}).

There are some member stars which should be noticed. In Figure \ref{fig: logr_Teff}(b), we can see that four member stars surrounded by a rectangle box have very high $\log R\rm'_{H\alpha}$ values, which belong to a same open cluster: NGC 2112 ($\log t = 9.315$). We check their spectra and find that they have very strong balmer emission lines. The open cluster is in the direction of the famous HII region known as Barnard's loop \citep{Haroon2017}. We suspect that the high $\log R\rm'_{H\alpha}$ values of this cluster are caused by interstellar medium. In (a) and (b) of Figure \ref{fig: logr_Teff}, some member stars have very low values so that they will pull down the mean value of $\log R'$ within an open cluster and increase the scatter obviously. So we exclude those stars whose $\log R\rm'_{CaK}$ $<$ -6.40 for the CaII K line and those stars whose $\log R\rm'_{H\alpha}$ $<$ -7.00 for the H$\alpha$ line.

\begin{figure}
\gridline{\fig{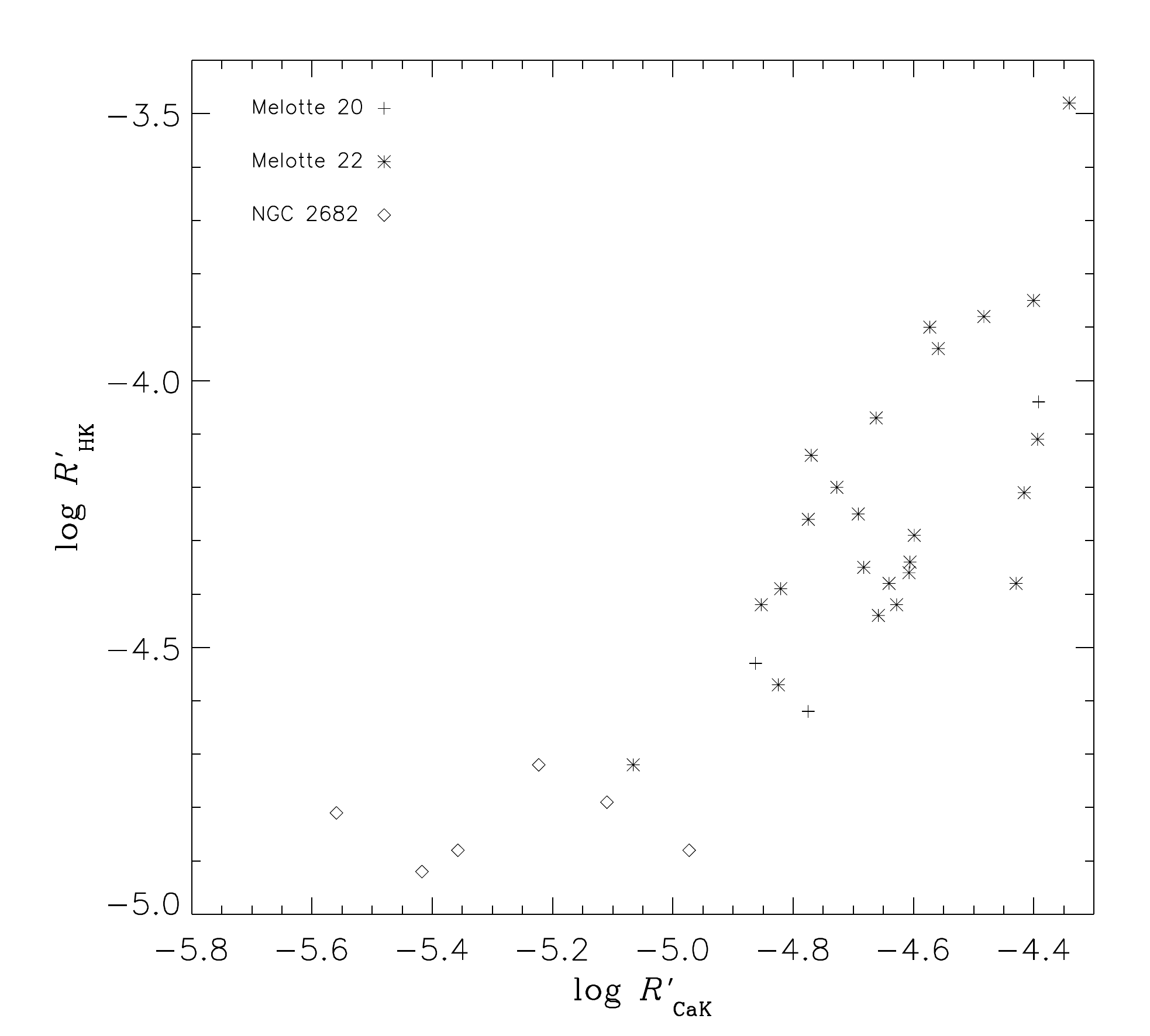}{0.52\textwidth}{(a)}
          \fig{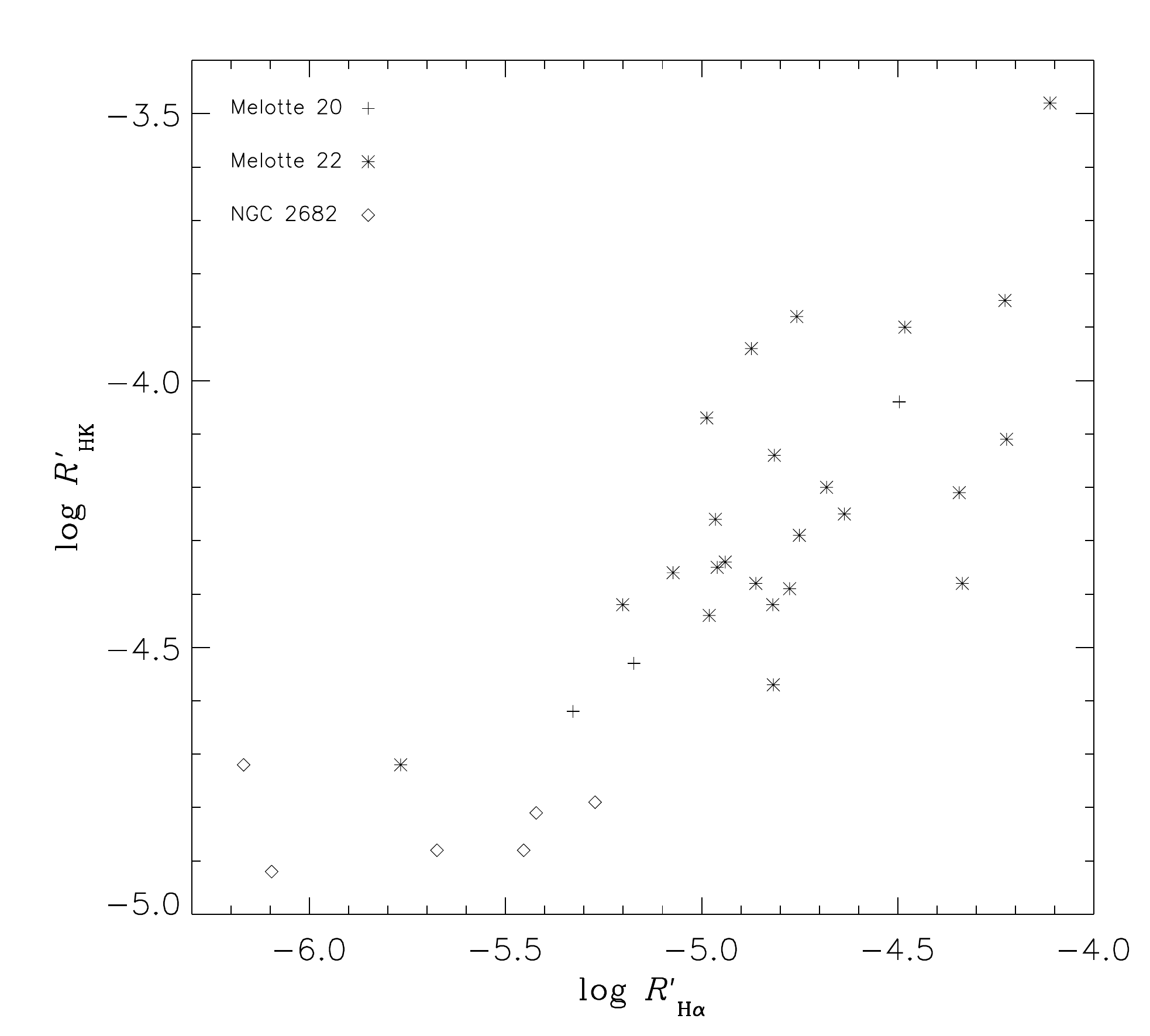}{0.52\textwidth}{(b)}
          }
\caption{The CA indices comparison for common stars between our sample and those from \cite{Mam2008}, which are member stars in three open clusters: Melotte 20, Melotte 22 and NGC 2682. The $\log R\rm'_{CaK}$ and $\log R\rm'_{H\alpha}$ are our CA indices. The $\log R\rm'_{HK}$ are from Table 5 of \cite{Mam2008}.   \label{fig:comparison with Mam08}}
\end{figure}

\cite{Mam2008} derived CA-age relation by using a traditional indicator $\log R'\rm_{HK}$ which was derived from S-values in the Mount Wilson HK project \citep{Vau1978,Noyes1984}. We cross match our results with Table 5 of \cite{Mam2008}. The comparison between our results and theirs is shown in Figure \ref{fig:comparison with Mam08}. The crossing match sample only includes three open clusters: Melotte 20, Melotte 22 and NGC 2682. From Figure \ref{fig:comparison with Mam08}, as $\log R\rm'_{CaK}$ or $\log R\rm'_{H\alpha}$ decreases, $\log R'\rm_{HK}$ also decreases. However, for those stars whose CA indices are low, our indices show a little larger scatter than theirs, which might be caused by different data processing methods. For stars whose $\rm EW$ are close to the basal lines, their $\rm EW'$ are close to zero and the $\log R\rm'_{CaK}$ and $\log R\rm'_{H\alpha}$ values discern more when taking the logarithm. In Appendix \ref{An illustration of logarithm effect}, we list a table (Table \ref{Table: Examples to show logarithm effect}) to illustrate it.

\subsection{$\log R'$ vs. $\log t$}
\subsubsection{$\log R'$ vs. $\log t$ in different $T\rm_{eff}$ ranges} \label{logR_logt_Teff}
\begin{figure}
\gridline{\fig{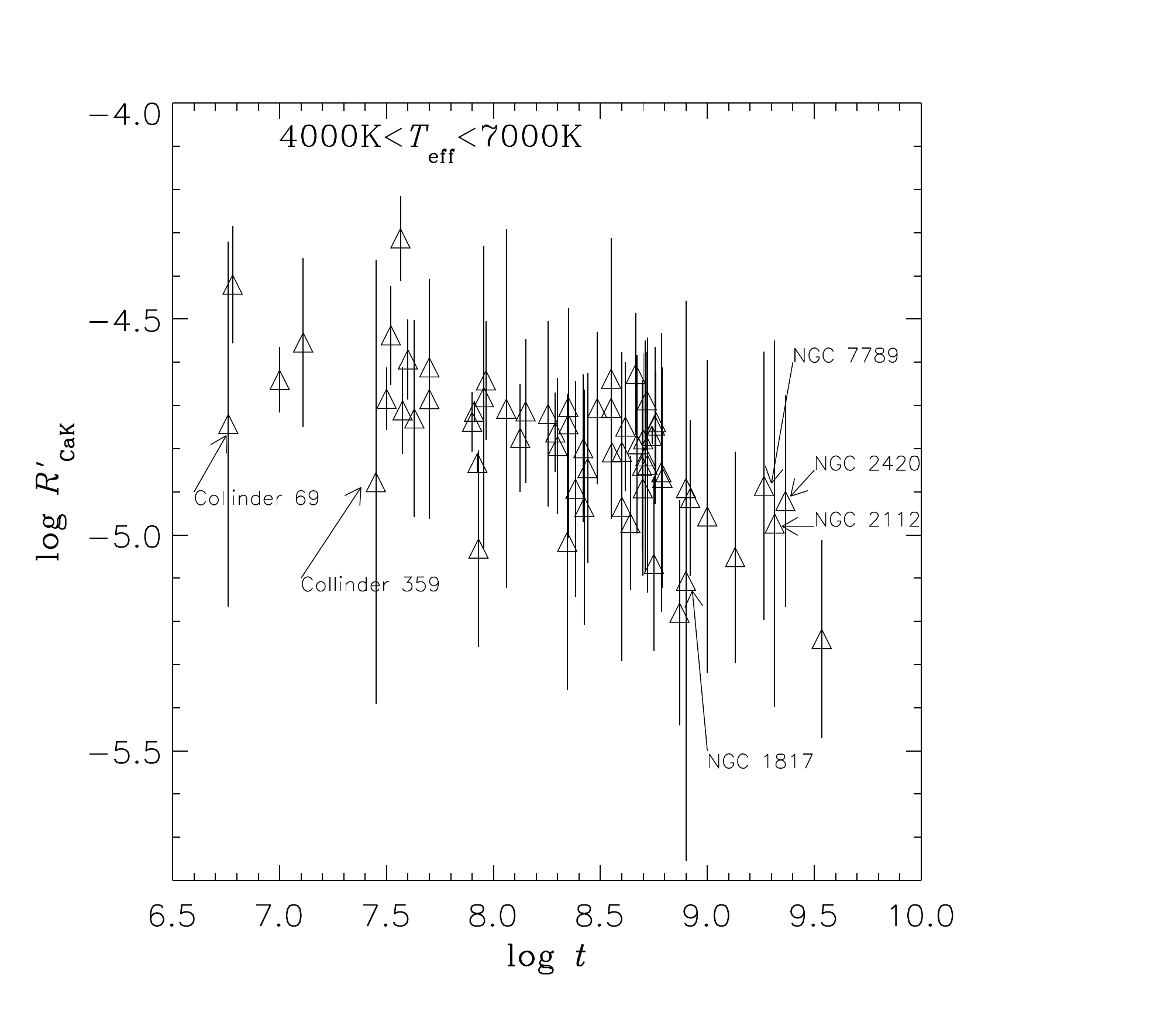}{0.5\textwidth}{(a)}
          \fig{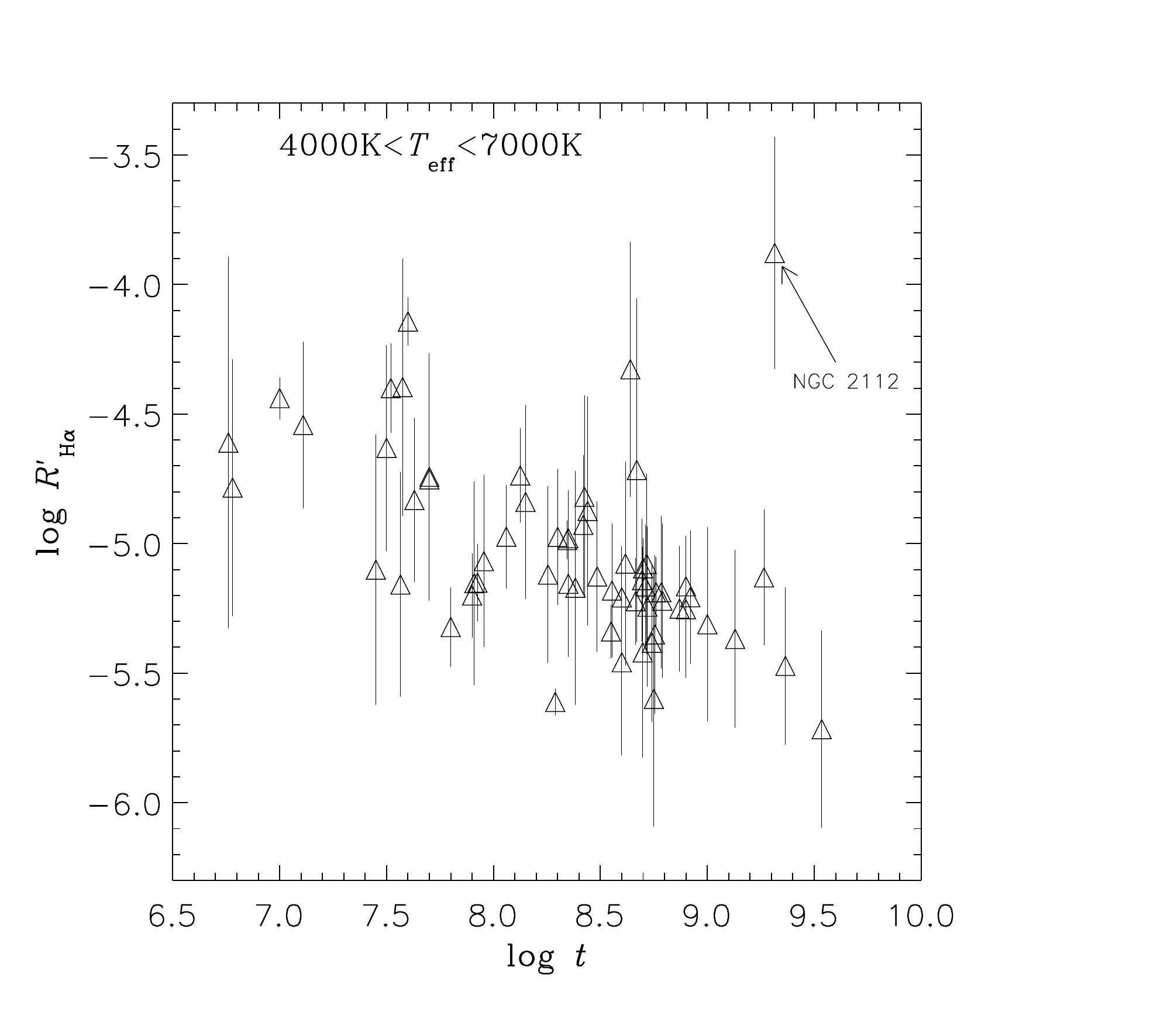}{0.5\textwidth}{(b)}
          }
\caption{The mean $\log R\rm'$ vs. age $\log t$ among open clusters. Each triangle represents a cluster and error bar indicates the standard deviation of the CA indices in each open cluster. Left-top corner gives the $T\rm_{eff}$ range of stars chosen to calculate the mean values. Those clusters with only one member star are not displayed. Arrows are used to specify the location of some open clusters.  \label{fig:logt_logr}}
\end{figure}

The mean value of $\log R\rm'_{CaK}$ ($\log R\rm'_{H\alpha}$) for each open cluster is calculated. Figure \ref{fig:logt_logr} plots this mean value vs. age of each cluster. The left-top corner gives the $T\rm_{eff}$ range of member stars chosen to calculate mean value. Clusters which have only one star are not displayed in this plot. From Figure \ref{fig:logt_logr} (a) we find that when stellar age $\log t$ $<$ 8.5 (0.3Gyr), the mean value of $\log R\rm'_{CaK}$ starts to decrease slowly as age increases. Then after $\log t=8.5$, the mean value decreases rapidly until $\log t = 9.53$ (3.4Gyr). \cite{Sod1991} pointed that the evolution of CA for a low-mass star may be going through three stages: a slow initial decline, a rapid decline at intermediate ages ($\sim$ 1-2 Gyr), and a slow decline for old stars like the sun. Although there are some differences on age ranges of each stage, our conclusion is consistent with that of \cite{Sod1991} for the two former stages. In our sample, the number of old open clusters ($\log t > 9.0$) is small and the age is only extended to $\log t = 9.53$, so it's hard to see whether there is a slow decline for old stars like the sun. From Figure \ref{fig:logt_logr} (b), the mean value of $\log R\rm'_{H\alpha}$ decreases from nearly $\log t=6.76$ (5.7Myr) to $\log t=9.53$ (3.4Gyr). Although it doesn't show the trend: a slow initial decline and then a rapid decline, we can see the same trend as CaII K if we divide $T\rm_{eff}$ range as done below.

From Figure \ref{fig:logt_logr}, we can see some open clusters deviate from the location of our expectation or have a large error bar. In panel(a), three old open clusters (NGC 7789, NGC 2112 and NGC 2420) show a little larger mean values. Their average [Fe/H] are relatively poor compared to young open clusters. For example, NGC 2420 has average [Fe/H] equal to -0.278$\pm$0.0462 (see Table \ref{clusters}). Poor [Fe/H] stars have relatively shallower metallic lines than rich [Fe/H] stars, leading to large $\rm EW_{CaK}$ and $\log R\rm'_{CaK}$, which may be the reason of these larger mean values. Some open clusters have one or two member stars whose $\log R\rm'_{CaK}$ values are very low, so that they pull down the mean values, like Collinder 69 and Collinder 359. We see that NGC 1817 has a very large error bar. This cluster has five member stars and all stars are with $T\rm_{eff}>$ 6500K. Three of them have large $\log R\rm'_{CaK}$ values, the other two have low $\log R\rm'_{CaK}$ values. The difference between the two groups is about 1 dex. In panel(b), NGC 2112 has a very large mean value. The reason is discussed above.

The scatter of $\log R'$ within an open cluster is large. In addition to measurement error, there are many other physical factors contributed to the scatter. Within an open cluster, different member star has different mass and rotation rate. Stellar mass and rotation rate can influence CA level \citep{Noyes1984,Mam2008}. Stellar cycle modulations also change CA level \citep{Baliunas1995,Lor2018}. Some stars in our sample may have flare or starspots, which affect CA level. Binaries and interstellar medium can also influence CA level. Appendix \ref{binaries and interstellar medium} simply discusses the impact of binaries and interstellar medium on $\log R'$. In our sample, some stars may not belong to open clusters and they affect the mean values. Besides, our data processing method also contributes to the scatter. For those stars whose $\rm EW$ are very close to the basal line, a small difference in $\rm EW$ between two stars can cause a large difference in $\log R'$ (see Table \ref{Table: Examples to show logarithm effect}).

\begin{figure}
\gridline{\fig{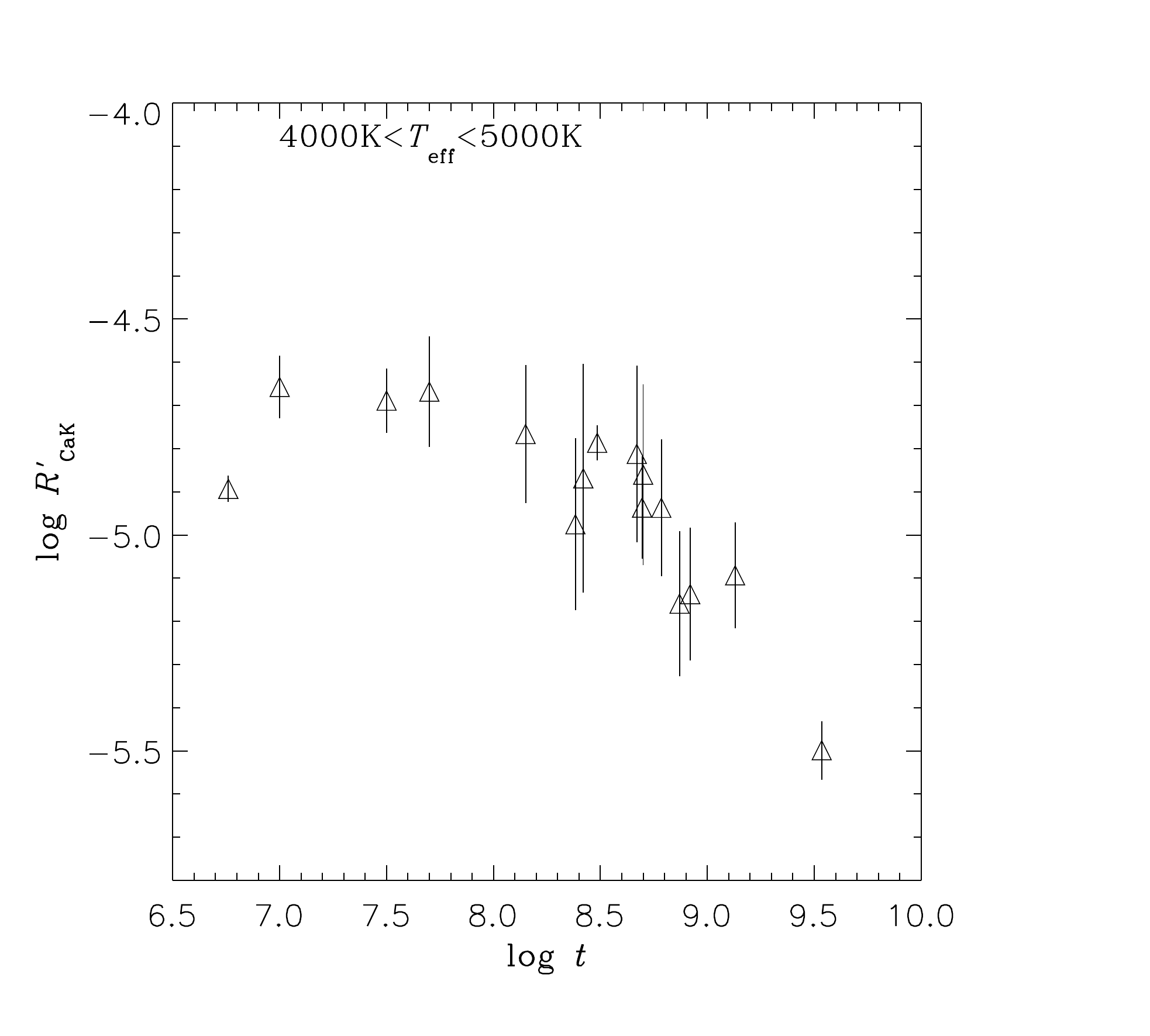}{0.4\textwidth}{(a)}
          \fig{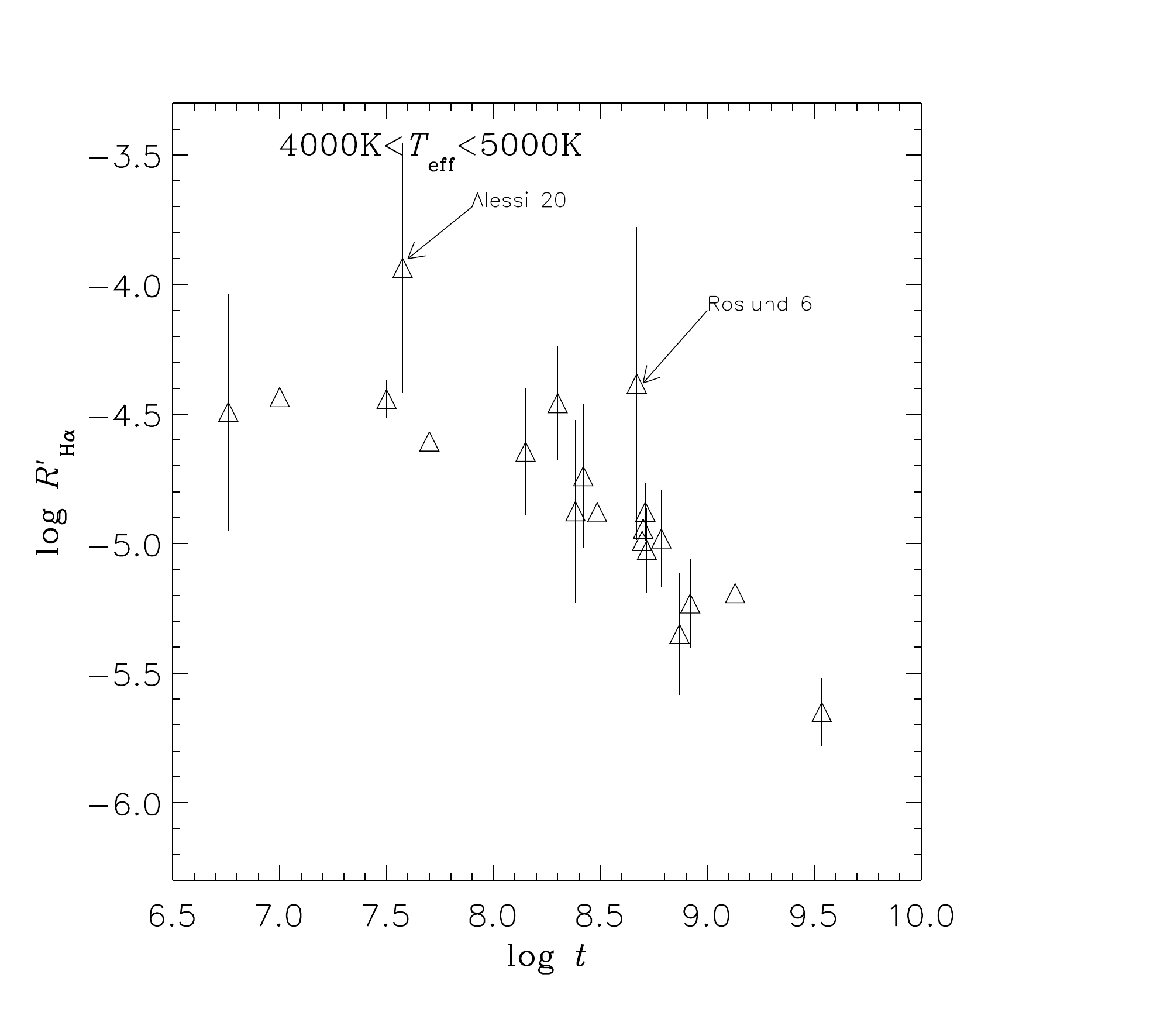}{0.4\textwidth}{(d)}
          }
\gridline{\fig{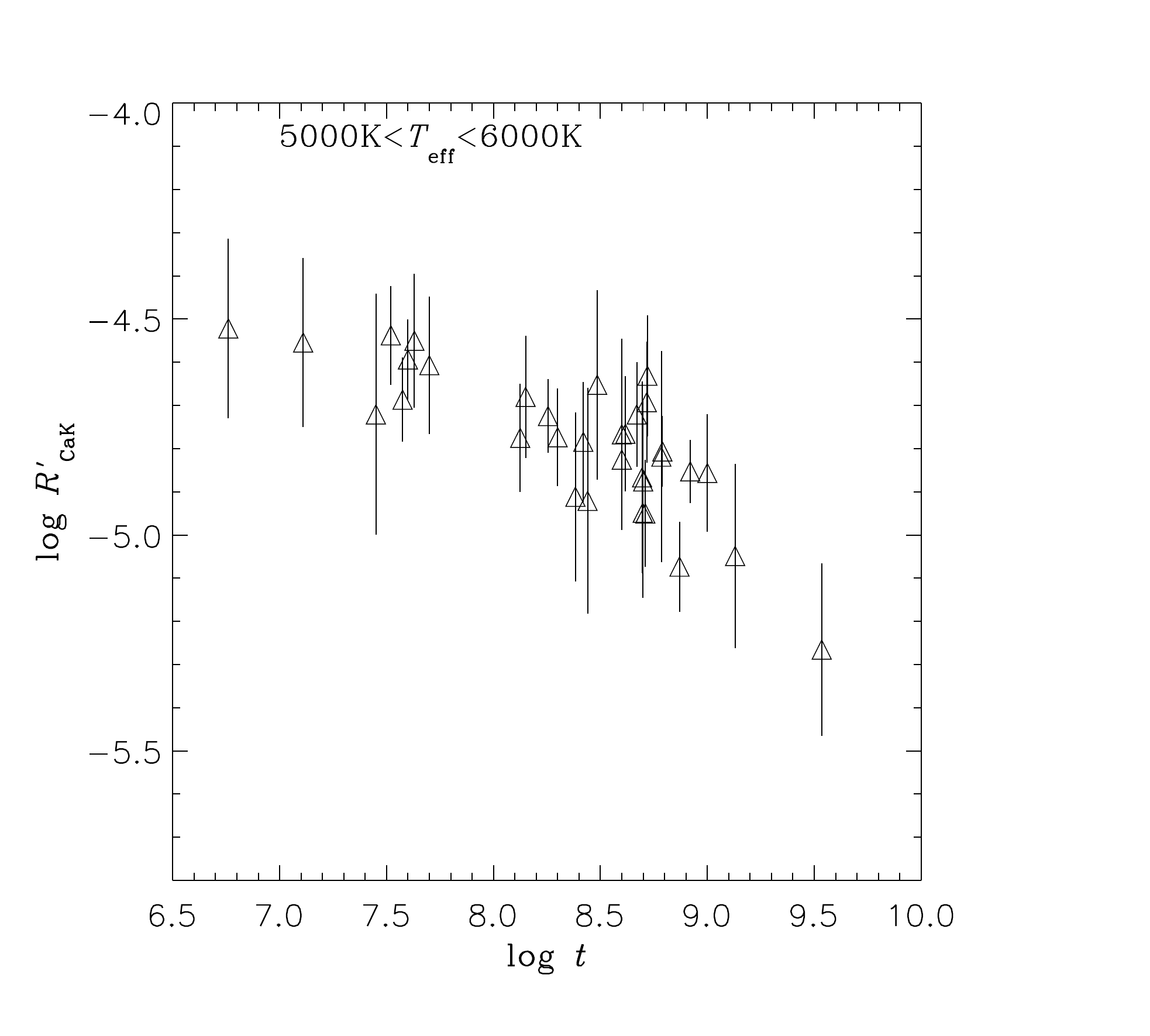}{0.4\textwidth}{(b)}
          \fig{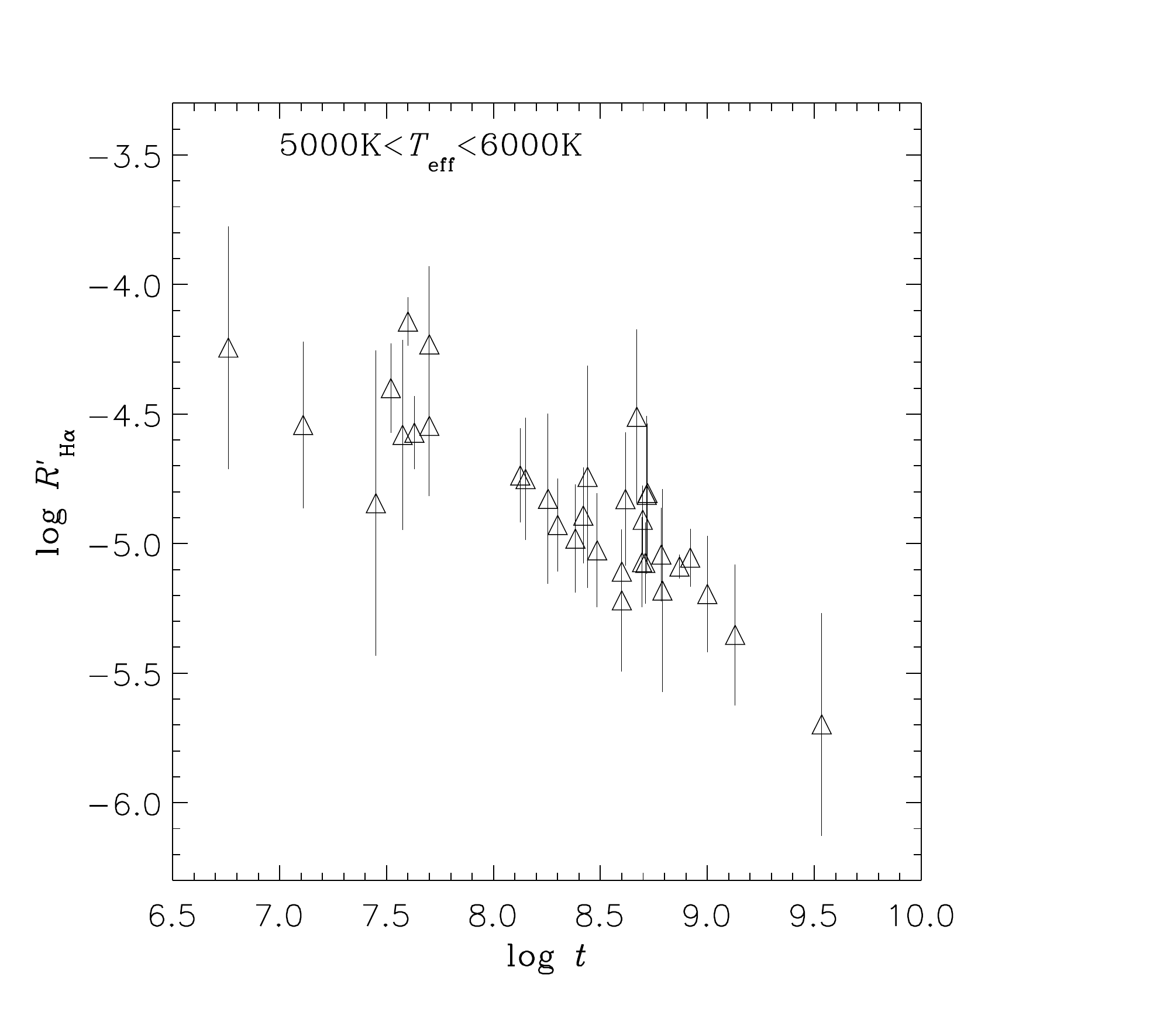}{0.4\textwidth}{(e)}
          }
\gridline{\fig{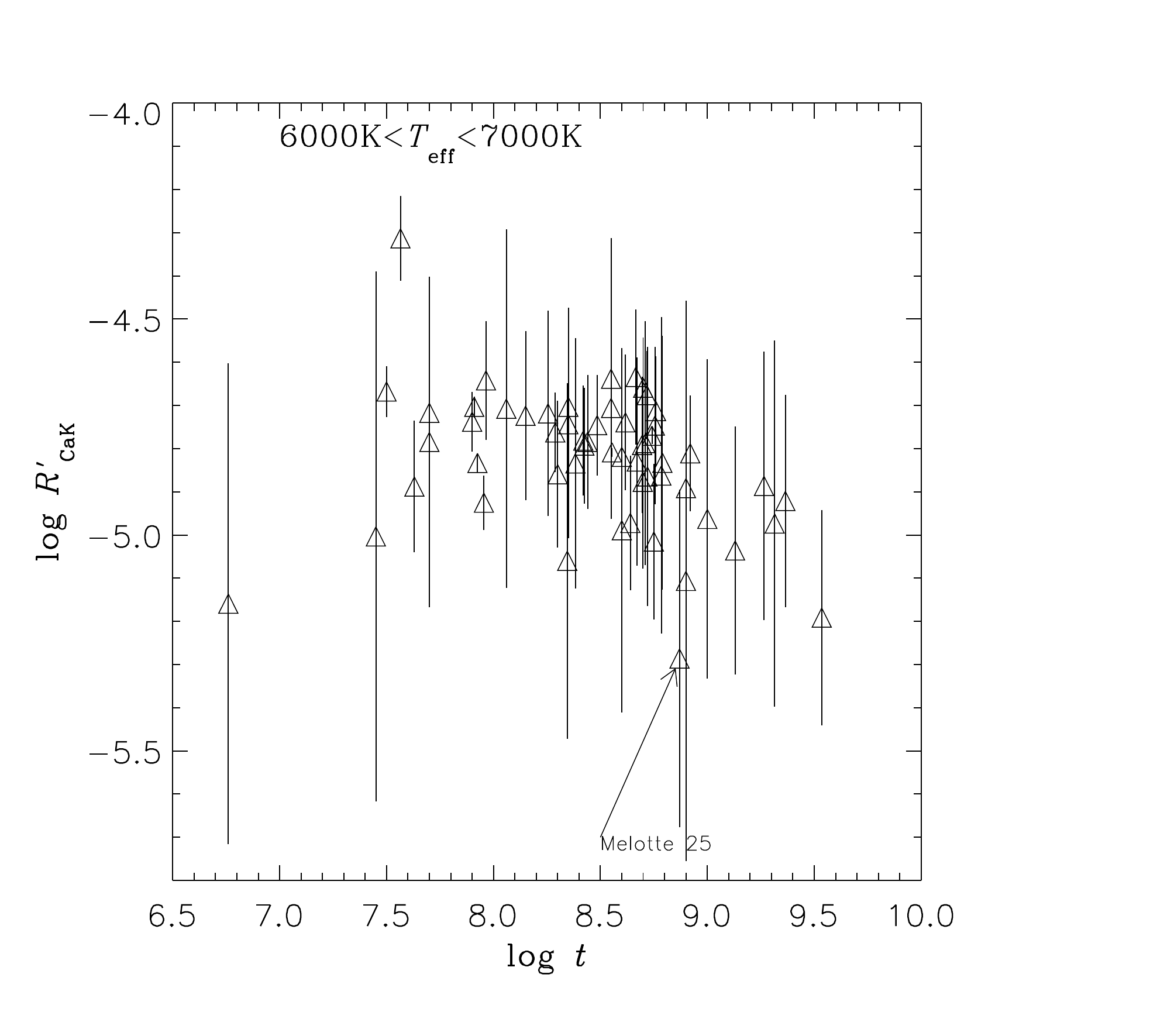}{0.4\textwidth}{(c)}
          \fig{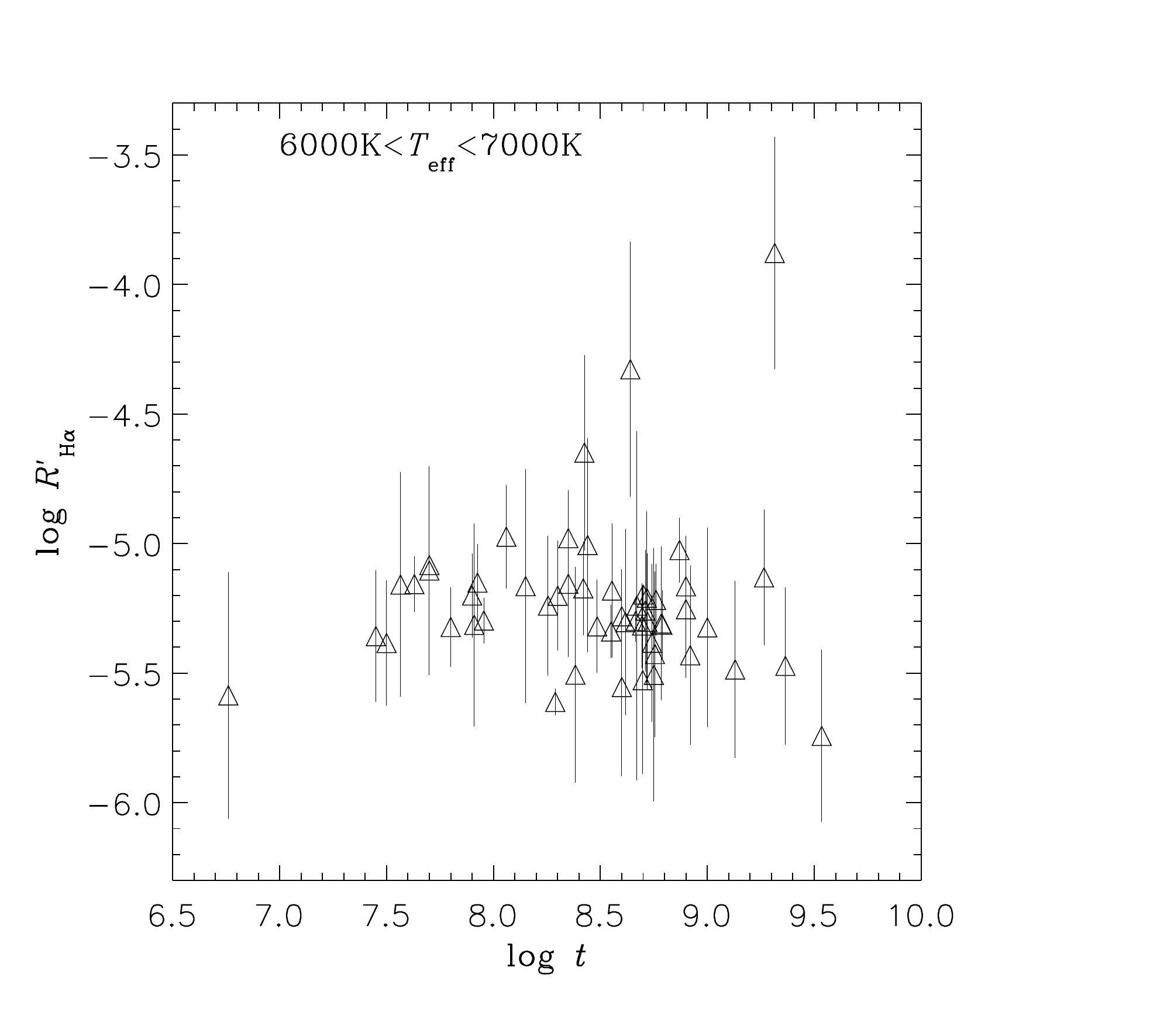}{0.4\textwidth}{(f)}
          }
\caption{The mean $\log R'$ vs. age ($\log t$) among open clusters in different $T\rm_{eff}$ ranges. Arrows are used to specify the location of some open clusters.  \label{fig:logt_logr1}}
\end{figure}

In order to decrease the influence of stellar mass on CA, we divide $T\rm_{eff}$ into three equal bins and plot the mean $\log R'$ vs. age $\log t$ again as shown in Figure \ref{fig:logt_logr1}. In all $T\rm_{eff}$ ranges, we can see the trend that as age increases the mean value decreases slowly or remains unchanged, and then decreases rapidly. The scatter is smaller at low $T\rm_{eff}$ range than at high $T\rm_{eff}$ range. That means $\log R'$ is more sensitive to stellar age at lower $T\rm_{eff}$, which is consistent with Figure 2 of \cite{ZhaoJK2011}. In their figure, the quantity $\log \rm S_{HK}$ used to indicate CA level discerned more from each other at redder color. The trend that CA shows a slow decline and then a rapid decline is more evident for cooler stars. This phenomena may be related to stellar inner structure. Those stars at low $T\rm_{eff}$ range have thicker convective zone than those at high $T\rm_{eff}$ range. So those stars at low $T\rm_{eff}$ range can maintain strong surface magnetic field at longer time scale than at high $T\rm_{eff}$ range \citep{FANG2018,West2008}.

There are some open clusters which deviate from locations of expectation or have a large error bar. Many of them are discussed above. In Figure \ref{fig:logt_logr1}(c), Melotte 25 has a low mean value and a large error bar. The reason is that this cluster has only three member stars at 6000K $<T\rm_{eff}<$ 7000K, of which one member star has a low $\log R\rm'_{CaK}$ value ($\log R\rm'_{CaK}=-5.81$), pulling down the mean value. In Figure \ref{fig:logt_logr1}(d), Alessi 20 has a larger mean value. With 4000K $<T\rm_{eff}<$ 5000K, this open cluster has only two member stars, whose [Fe/H] are poorer compared to the other member stars. One of these two stars shows very strong H$\alpha$ emission line. Maybe these two stars are not member stars of the cluster. Roslund 6 has only two member stars with 4000K $<T\rm_{eff}<$ 5000K. The $\rm EW_{H\alpha}$ of these two stars are very large. One is 15.13\AA, the other is 3.90\AA. Their spectra show very strong H$\alpha$ emission line. Not just H$\alpha$, there are other emission lines in these two spectra such as: H$\beta$, CaII HK, NII and so on. Maybe the two stars are in a special term. For example, they have large spots on stellar surface.

\subsubsection{$\log R'$ vs. $\log t$ in a narrow $\rm [Fe/H]$ range}

\begin{figure}
\gridline{\fig{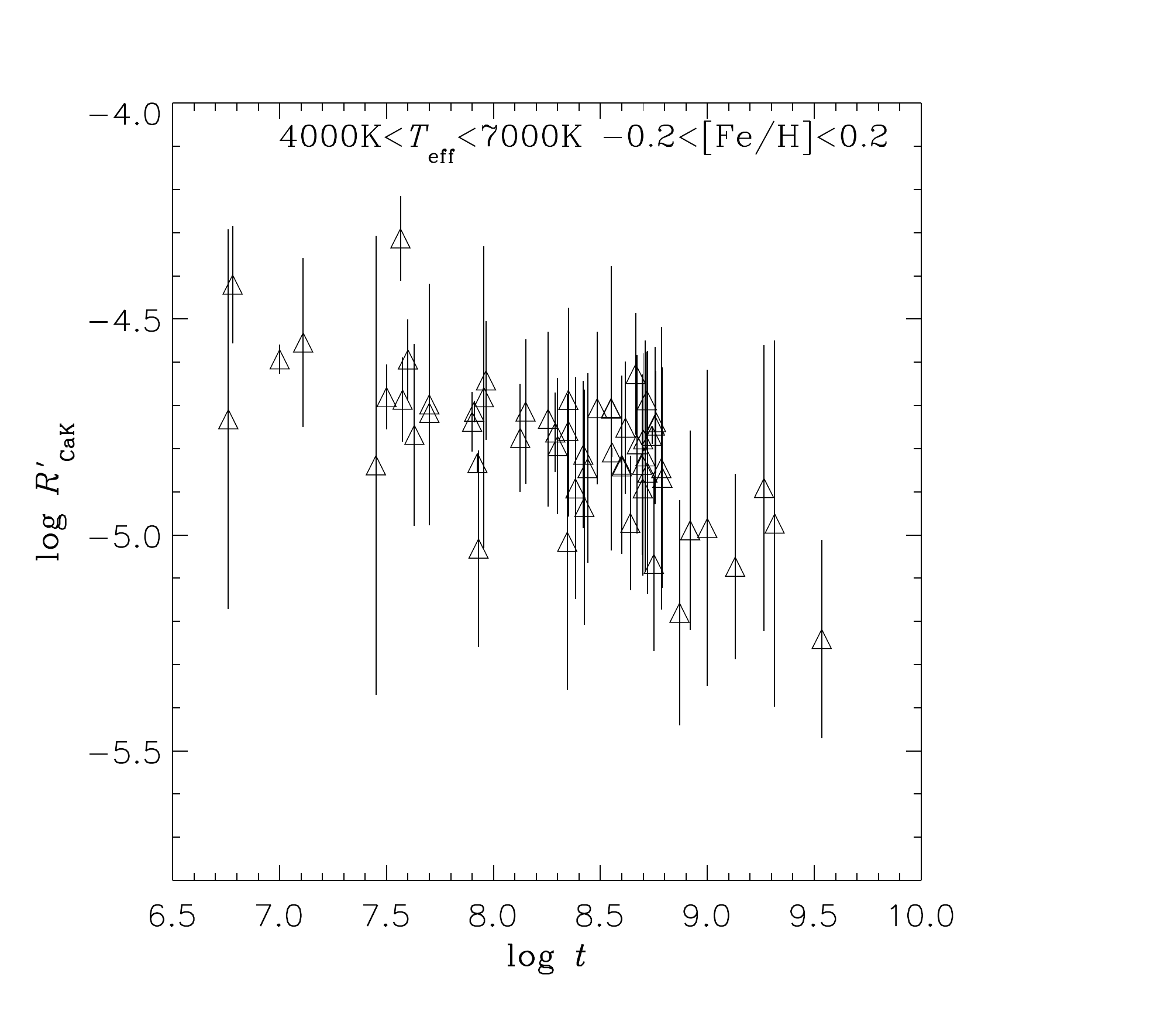}{0.5\textwidth}{(a)}
          \fig{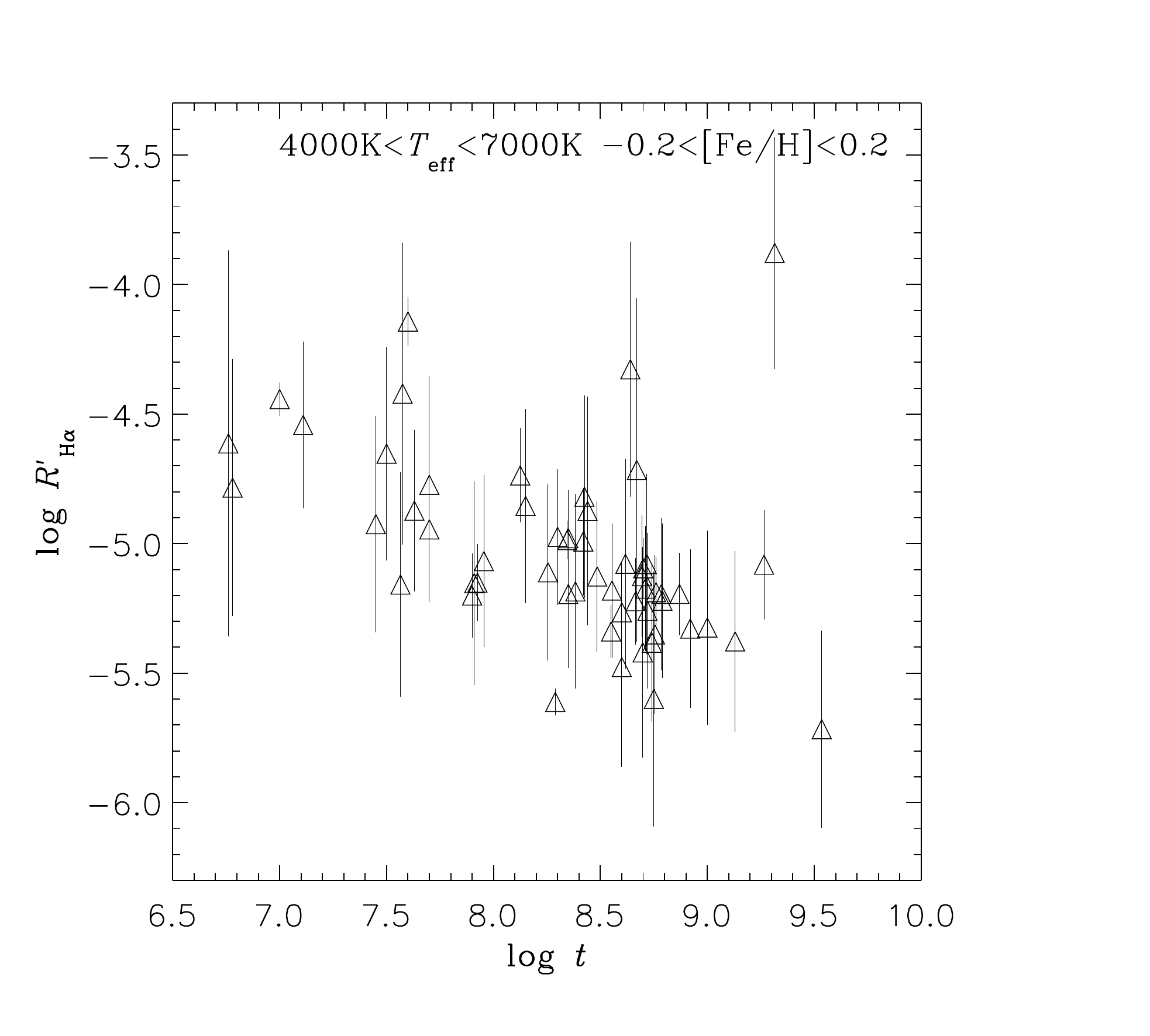}{0.5\textwidth}{(b)}
          }
\caption{The mean $\log R'$ vs. age ($\log t$) among open clusters with 4000K $<T\rm_{eff}<$ 7000K and $-0.2<$ [Fe/H] $<0.2$.    \label{fig:logt_logr_feh}}
\end{figure}

[Fe/H] has more effect on $\log R\rm'_{CaK}$ than $\log R\rm'_{H\alpha}$ \citep{Roc1998,Roc2000,Lor2016}. Maybe there is a negative correlation between $\log R\rm'_{CaK}$ and [Fe/H]. We narrow [Fe/H] range to $-0.2<$ [Fe/H] $<0.2$ and plot the mean $\log R'$ vs. age $\log t$ again. The sample is splited on a star-by-star basis. Figure \ref{fig:logt_logr_feh} shows $\log R'$ vs. $\log t$ with 4000K $<T\rm_{eff}<$ 7000K and $-0.2<$ [Fe/H] $<0.2$. By comparing Figure \ref{fig:logt_logr_feh} and Figure \ref{fig:logt_logr}, we find that there is no obvious difference. Figure \ref{fig:logt_logr2}(a) and (c) shows $\log R'$ vs. $\log t$ with 4000K $<T\rm_{eff}<$ 5500K and without [Fe/H] limit. Figure \ref{fig:logt_logr2}(b) and (d) shows $\log R'$ vs. $\log t$ with 4000K $<T\rm_{eff}<$ 5500K and $-0.2<$ [Fe/H] $<0.2$. By comparsion, no obvious difference is formed. We also see no large difference of $\log R'$ vs. $\log t$ relation when narrowing [Fe/H] range to $-0.1<$ [Fe/H] $<0.1$.

\subsection{fitting between $\log R'$ and $\log t$ at low $T\rm_{eff}$ range}
\begin{figure}
\gridline{\fig{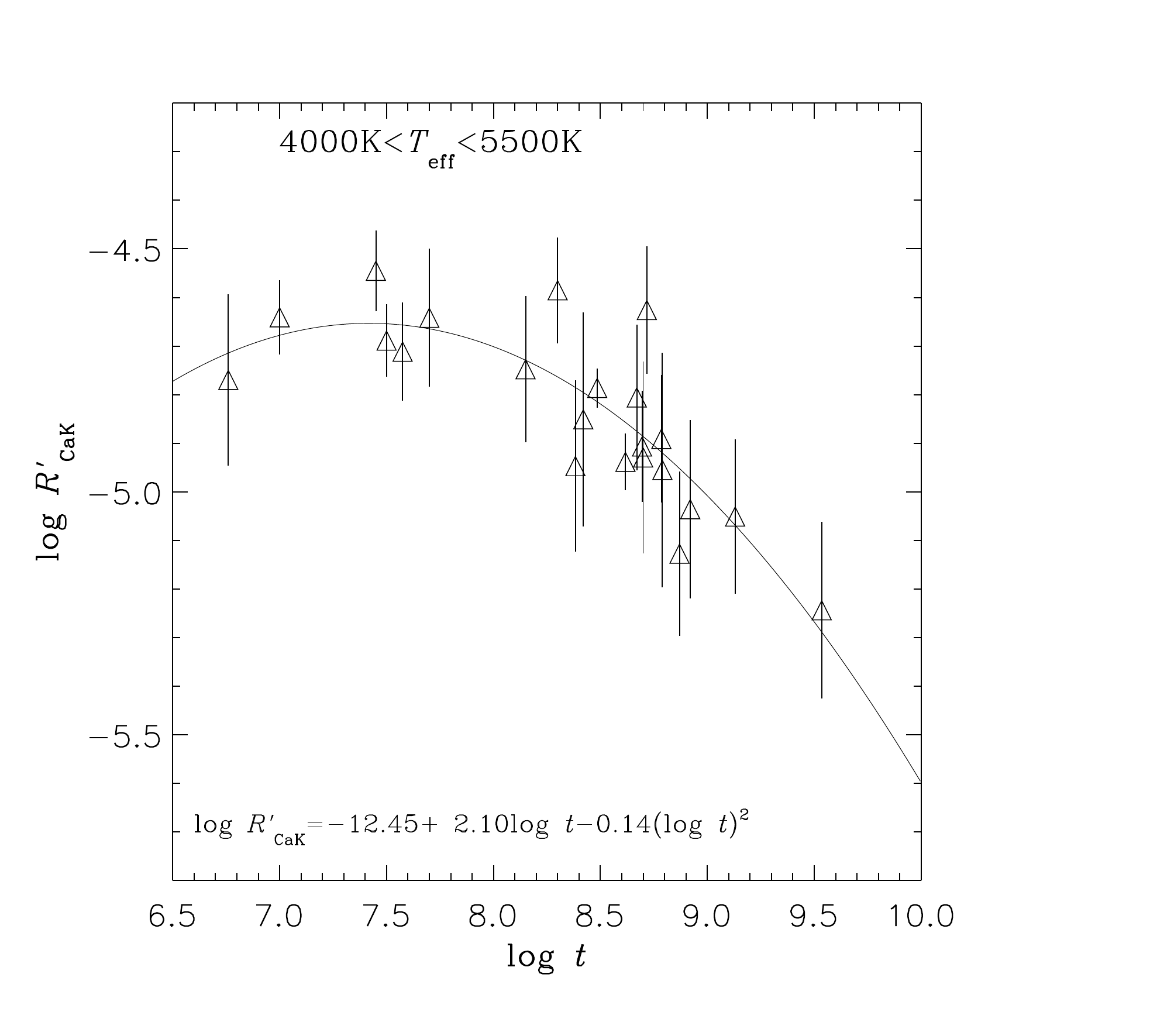}{0.5\textwidth}{(a)}
          \fig{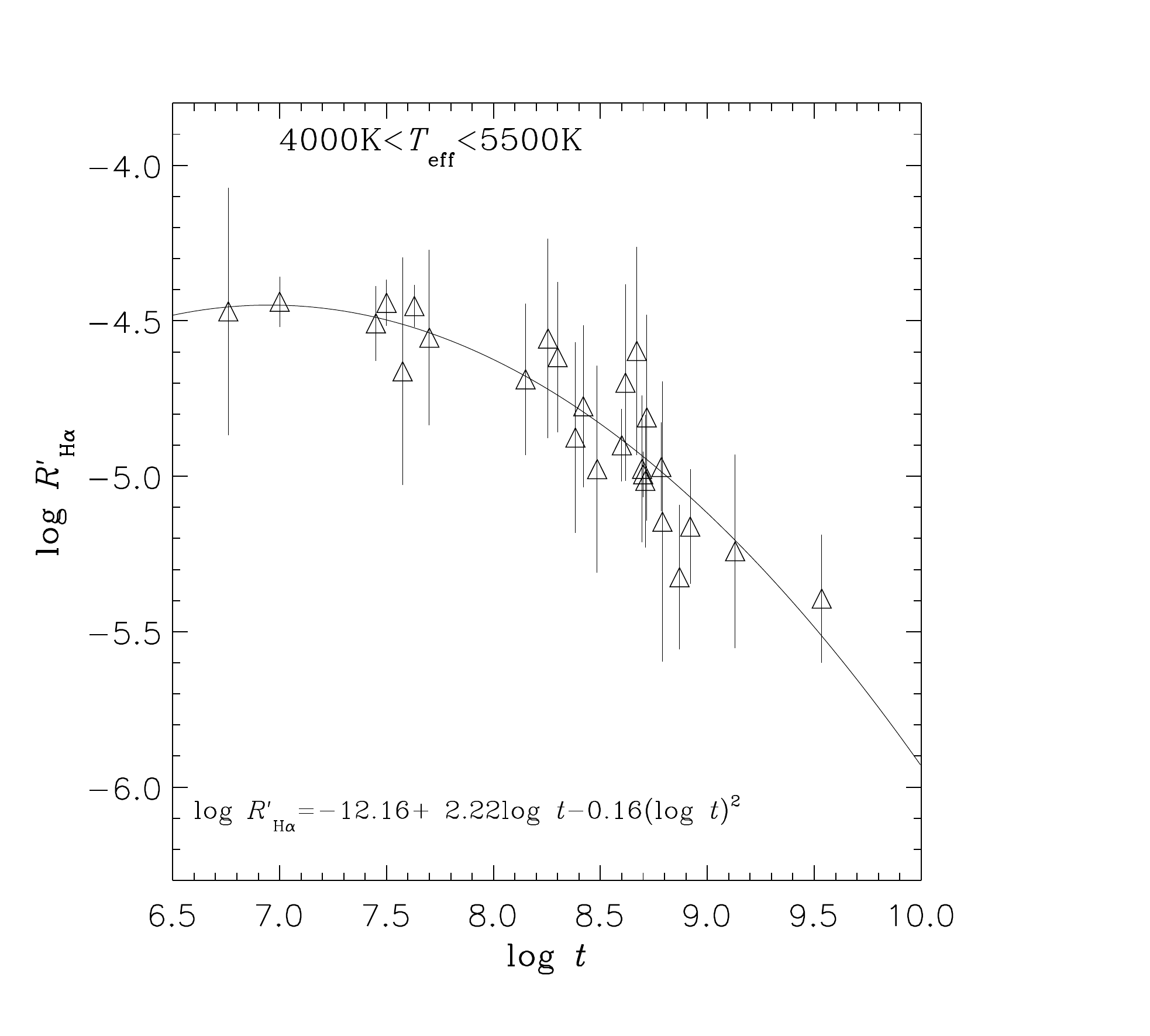}{0.5\textwidth}{(c)}
          }
\gridline{\fig{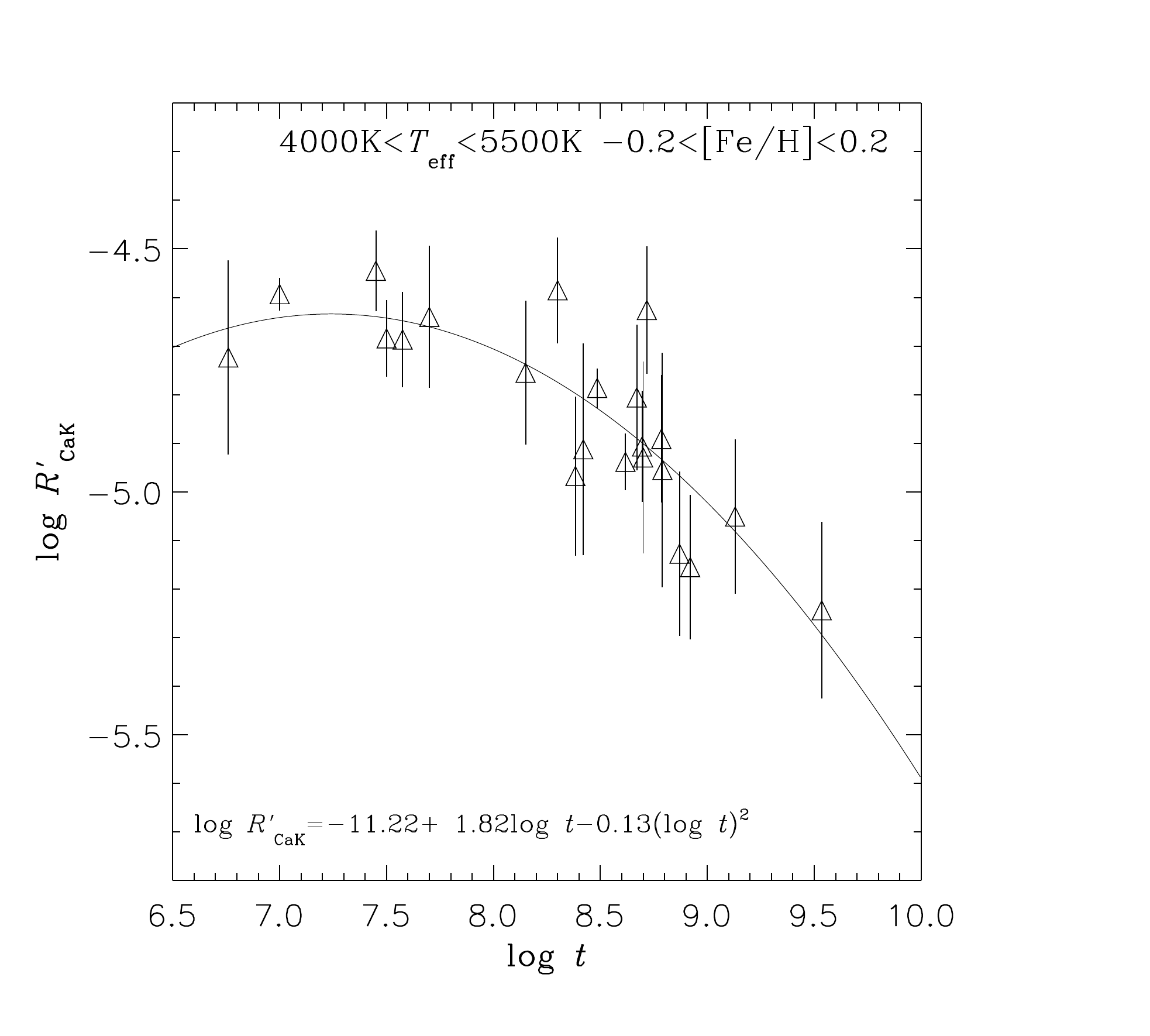}{0.5\textwidth}{(b)}
          \fig{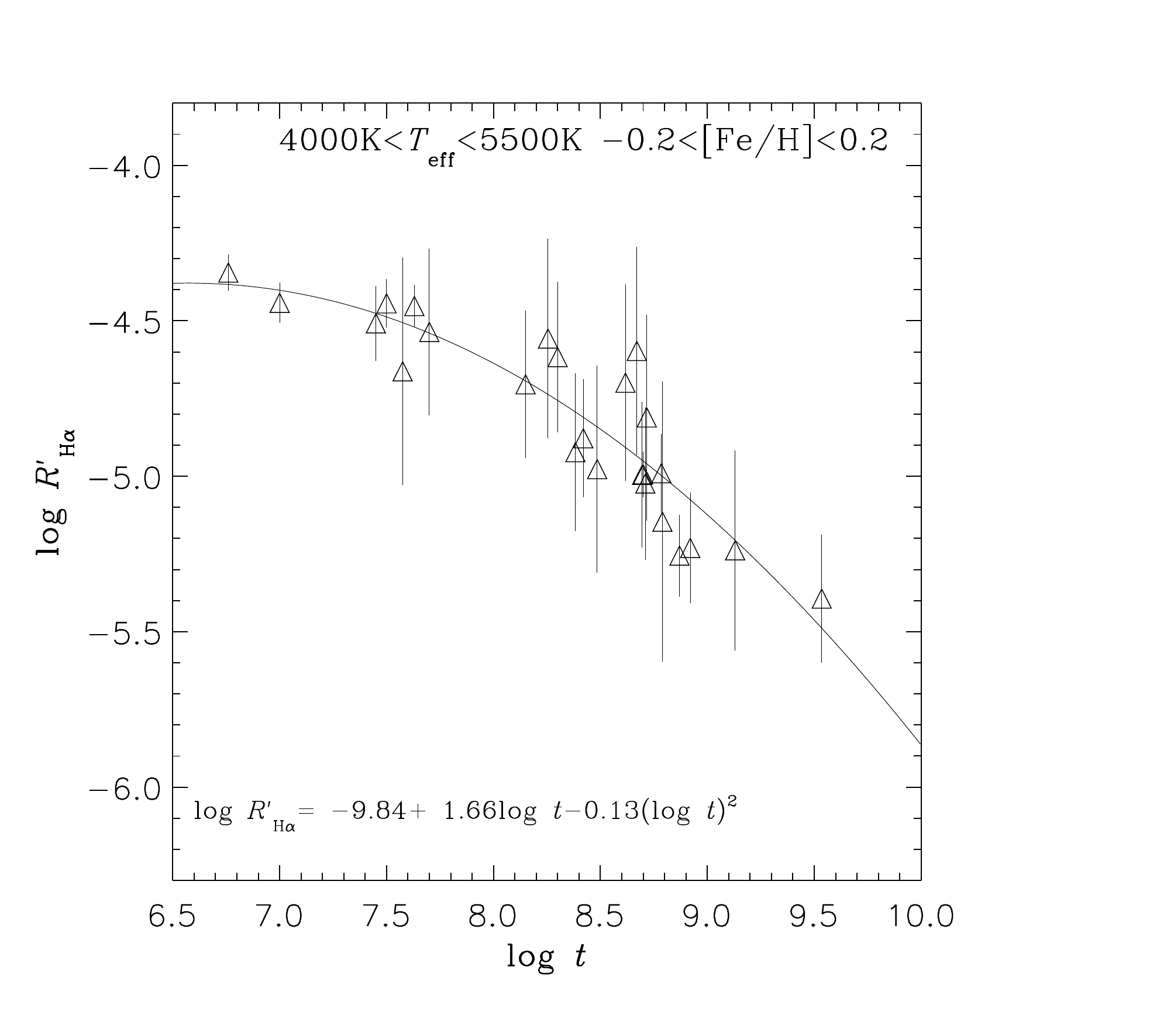}{0.5\textwidth}{(d)}
          }
\caption{The mean $\log R'$ vs. age ($\log t$) with 4000K $<T\rm_{eff}<$ 5500K in two [Fe/H] ranges. Quadratic function is used to fit these data points. Relationship is listed in the left-bottom corner. For H$\alpha$, two stars of Alessi 20 and two stars of Roslund 6 mentioned in \ref{logR_logt_Teff} are removed.   \label{fig:logt_logr2}}
\end{figure}

Quadratic function is used to fit data points with 4000K $<T\rm_{eff}<$ 5500K in two [Fe/H] ranges. Figure \ref{fig:logt_logr2} shows fitting curves and relationships. For H$\alpha$, two stars of Alessi 20 and two stars of Roslund 6 mentioned in \ref{logR_logt_Teff} are removed. The relationships are also listed in Equations \ref{Equation: lorcak_logt}-\ref{Equation: lorha_logt_m02p02}. Age for field stars can be approximately estimated by solving these quadratic equations. Via Monte Carlo simulation, a distribution of $\log t$ can be obtained with a $\log R'$ value and its error. For Equations \ref{Equation: lorcak_logt} and \ref{Equation: lorcak_logt_m02p02}, we calculate two distribution of $\log t$ at two $\log R\rm'_{CaK}$ values ($\log R\rm'_{CaK}=-4.90, -5.23$). The error of $\log R\rm'_{CaK}$ is set to 0.15dex. The error of $\log t$  are about 0.40dex, 0.28dex at $\log t=8.75, 9.44$ corresponding to $\log R\rm'_{CaK}=-4.90, -5.23$. The error of $\log t$ at $\log t=9.44$ is smaller than at $\log t=8.75$. This is because the fitting curve gets steeper when $\log t$ increases and $\log R\rm'_{CaK}$ is projected to a smaller range of $\log t$.  For Equations \ref{Equation: lorha_logt} and \ref{Equation: lorha_logt_m02p02}, the error of $\log t$ are about 0.40dex, 0.28dex at $\log t=8.60, 9.40$ corresponding to $\log R\rm'_{H\alpha}=-4.90, -5.40$. The error of $\log R\rm'_{H\alpha}$ is set to 0.20dex.

Equations \ref{Equation: lorcak_logt} and \ref{Equation: lorha_logt} are used to estimate ages of corresponding clusters whose $\log t>8.00$. The results and relative error are shown in Table \ref{Estimated ages of open clusters}. The accuracy of Equation \ref{Equation: lorcak_logt} is about 40\%, while the accuracy of Equation \ref{Equation: lorha_logt} is about 60\%. The ages of NGC 1647 and Ascc 10 can't be estimated by Equation \ref{Equation: lorcak_logt} because the two clusters have very large $\log R\rm'_{CaK}$ mean values which exceed the maximum value of Equation \ref{Equation: lorcak_logt}.

Equation \ref{Equation: lorcak_logt} and Equation \ref{Equation: lorha_logt} are also used to estimate ages of open clusters whose ages are not found in literatures. However, only the open cluster RSG 1 is available to estimate age. The cluster has $\log t = 8.69$ estimated by Equation \ref{Equation: lorcak_logt} and $\log t = 8.52$ estimated by Equation \ref{Equation: lorha_logt}. In a following paper, we will use Equations \ref{Equation: lorcak_logt}-\ref{Equation: lorha_logt_m02p02} to roughly estimate ages of field stars.

\begin{equation}\label{Equation: lorcak_logt}
\log R'_{\rm CaK}=-12.45+2.10\log t-0.14(\log t)^2,
\end{equation}
\begin{equation}\label{Equation: lorcak_logt_m02p02}
\log R'_{\rm CaK}=-11.22+1.82\log t-0.13(\log t)^2,   \rm -0.2<[Fe/H]<0.2
\end{equation}
\begin{equation}\label{Equation: lorha_logt}
\log R'_{\rm H\alpha}=-12.16+2.22\log t-0.16(\log t)^2,
\end{equation}
\begin{equation}\label{Equation: lorha_logt_m02p02}
\log R'_{\rm H\alpha}=-9.84+1.66\log t-0.13(\log t)^2,   \rm -0.2<[Fe/H]<0.2
\end{equation}

\startlongtable
\begin{deluxetable}{cccccc}
\tablecaption{Estimated ages of corresponding open clusters whose $\log t>8.00$\label{Estimated ages of open clusters}}
\tablehead{
\colhead{name} & \colhead{$t$}  & \colhead{Equation \ref{Equation: lorcak_logt}}  & \colhead{relative error}  & \colhead{Equation \ref{Equation: lorha_logt}}  & \colhead{relative error}\\
\colhead{} & \colhead{Myr}  & \colhead{Myr}  & \colhead{}  & \colhead{Myr}  & \colhead{}
}
\startdata
  Melotte\_22 & 141 & 176 & 25\% & 147 & 4\%\\
  NGC\_2168 & 180 &  &  & 58 & 68\%\\
  NGC\_1647 & 200 &  &  & 93 & 53\%\\
  NGC\_1039 & 242 & 732 & 203\% & 380 & 57\%\\
  Stock\_10 & 263 & 405 & 54\% & 237 & 10\%\\
  ASCC\_23 & 305 & 249 & 19\% & 582 & 91\%\\
  NGC\_1342 & 398 &  &  & 424 & 6\%\\
  NGC\_1750 & 414 & 696 & 68\% & 157 & 62\%\\
  Roslund\_6 & 468 & 292 & 38\% & 80 & 83\%\\
  NGC\_1662 & 495 & 580 & 17\% & 580 & 17\%\\
  ASCC\_41 & 501 & 661 & 32\% & 623 & 24\%\\
  Collinder\_350 & 513 &  &  & 676 & 32\%\\
  ASCC\_10 & 521 &  &  & 283 & 46\%\\
  NGC\_2281 & 610 & 526 & 14\% & 566 & 7\%\\
  IC\_4756 & 617 & 766 & 24\% & 1088 & 76\%\\
  Melotte\_25 & 741 & 1788 & 141\% & 1950 & 163\%\\
  NGC\_2632 & 832 & 1169 & 41\% & 1150 & 38\%\\
  NGC\_752 & 1349 & 1258 & 7\% & 1495 & 11\%\\
  NGC\_2682 & 3428 & 2918 & 15\% & 2403 & 30\%\\
\enddata
\tablecomments{The first column is the names of open clusters. The second column is the ages ($t$) of clusters from references. The third and fourth columns are the estimated ages from Equation \ref{Equation: lorcak_logt} and its relative error. The fifth and sixth columns are the estimated ages from Equation \ref{Equation: lorha_logt} and its relative error.}
\end{deluxetable}

\section{Conclusion and outlook} \label{sec:conclusion}

In this paper, we investigate the CA-age relationship by using the largest sample of open clusters in the LAMOST survey. Fang's method \citep{FANG2018} is used to calculate excess fractional luminosity $\log R\rm'_{CaK}$, $\log R\rm'_{H\alpha}$ of every member star which can be used to indicate CA level. In this method, we use 10\% quantile in $\rm EW$ to obtain the basal lines. Excess equivalent width $\rm EW'$ can be obtained after subtracting $\rm EW_{basal}$. Then $R'$ can be obtained via $R'\rm=EW'\times \chi$, of which $\chi$ is the ratio of the surface continuum flux near the line to the stellar surface bolometric flux from model spectra.

For each open cluster, the average $\log R\rm'_{CaK}$, $\log R\rm'_{H\alpha}$ can be calculated. For CaII K, 1091 member stars of 82 open clusters have $\log R'\rm_{CaK}$ measurements. For H$\alpha$, 1118 member stars of 83 open clusters have $\log R'\rm_{H\alpha}$ measurements. Then the relationship between the average $\log R'$ and the age can be studied in different $T\rm_{eff}$ ranges and [Fe/H] ranges. We find that CA starts to decrease slowly from $\log t=6.70$ to $\log t=8.50$, then decreases rapidly until $\log t=9.53$, which is consistent with the point of \cite{Sod1991}. The trend is more evident for cooler stars. This phenomena may be related to stellar inner structure. Compared to stars at high $T\rm_{eff}$, stars at low $T\rm_{eff}$ have thicker convective zone, such that they can maintain strong surface magnetic field at longer time scale. We narrow [Fe/H] range to $-0.2<$ [Fe/H] $<0.2$ and find that there is no obvious difference. Finally, we construct quadratic functions between $\log R'$ and $\log t$ with 4000K $<T\rm_{eff}<$ 5500K, which can be used to roughly estimate ages of field stars with accuracy about 40\% for $\log R'\rm_{CaK}$ and 60\% for $\log R'\rm_{H\alpha}$.

The LAMOST telescope has obtained about 9 million spectra. The relations shown in Equations \ref{Equation: lorcak_logt}-\ref{Equation: lorha_logt_m02p02} suggest that $\log R'$ can be used to roughly estimate stellar ages for dwarfs. With reliable stellar ages, the evolution of the thin disk can be investigated. For example, we can study the spatial age distribution and relations between the stellar age and velocity. Older open clusters are needed to extend the CA-age relation. Medium resolution spectra (R $\sim$ 8000) being obtained with the ongoing LAMOST survey may improve the CA-age relation in the near future.

\acknowledgments

This work is supported by the Astronomical Big Data Joint Research Center, co-founded by the National Astronomical Observatories, Chinese Academy of Sciences and the Alibaba Cloud, the National Natural Science Foundation of China under grant No.11973048, 11573035, 11625313, 11890694.  Support from the US National Science Foundation (AST-1358787) to Embry-Riddle Aeronautical University is acknowledged. Guoshoujing Telescope (the Large Sky Area Multi-Object Fiber Spectroscopic Telescope LAMOST) is a National Major Scientific Project built by the Chinese Academy of Sciences. Funding for the project has been provided by the National Development and Reform Commission. LAMOST is operated and managed by the National Astronomical Observatories, Chinese Academy of Sciences.

\appendix
\section{Measurement error of $\rm EW$ and $\log R'$}\label{Measurement error of logR}

\begin{figure}
\gridline{\fig{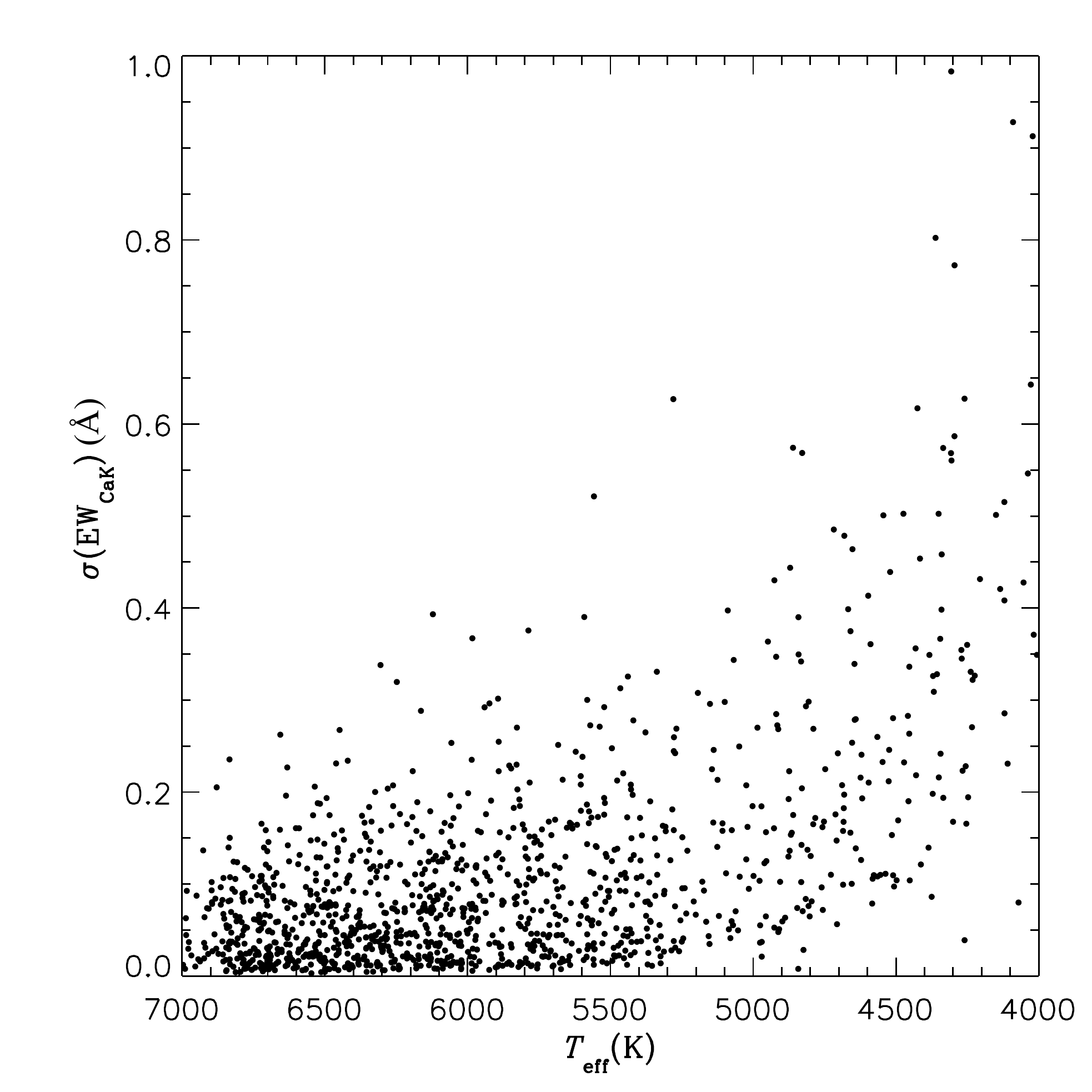}{0.5\textwidth}{(a)}
          \fig{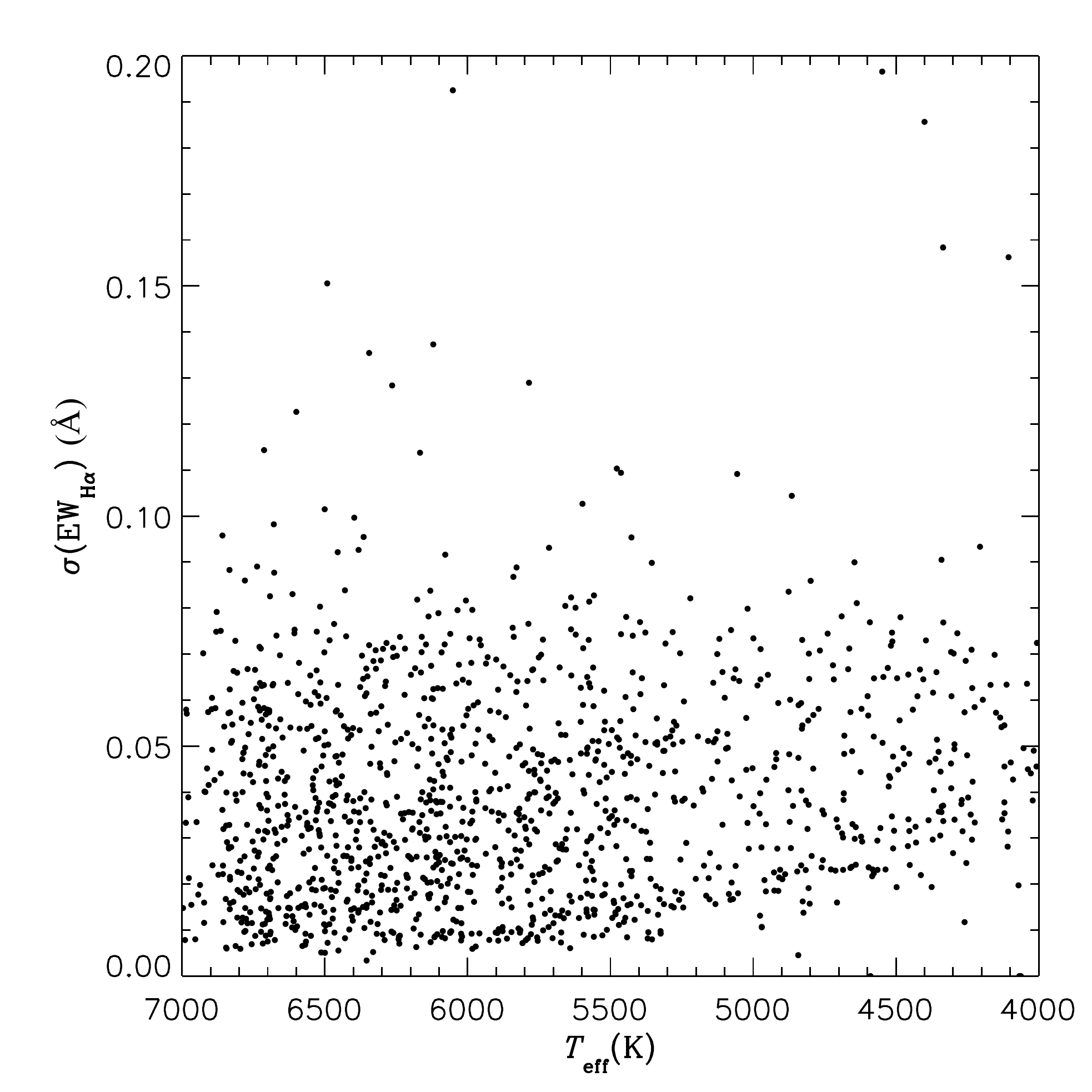}{0.5\textwidth}{(b)}
          }
\gridline{\fig{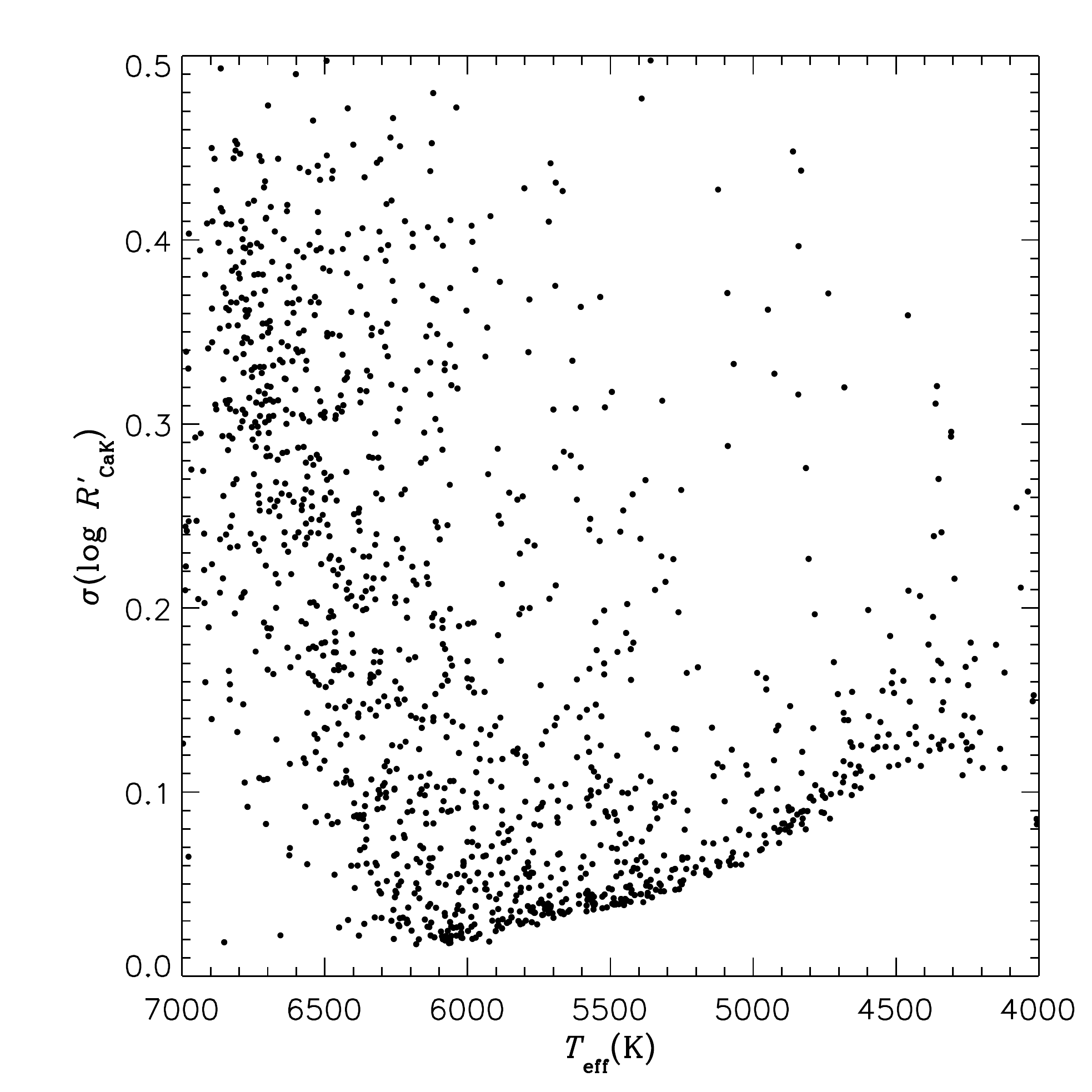}{0.5\textwidth}{(c)}
          \fig{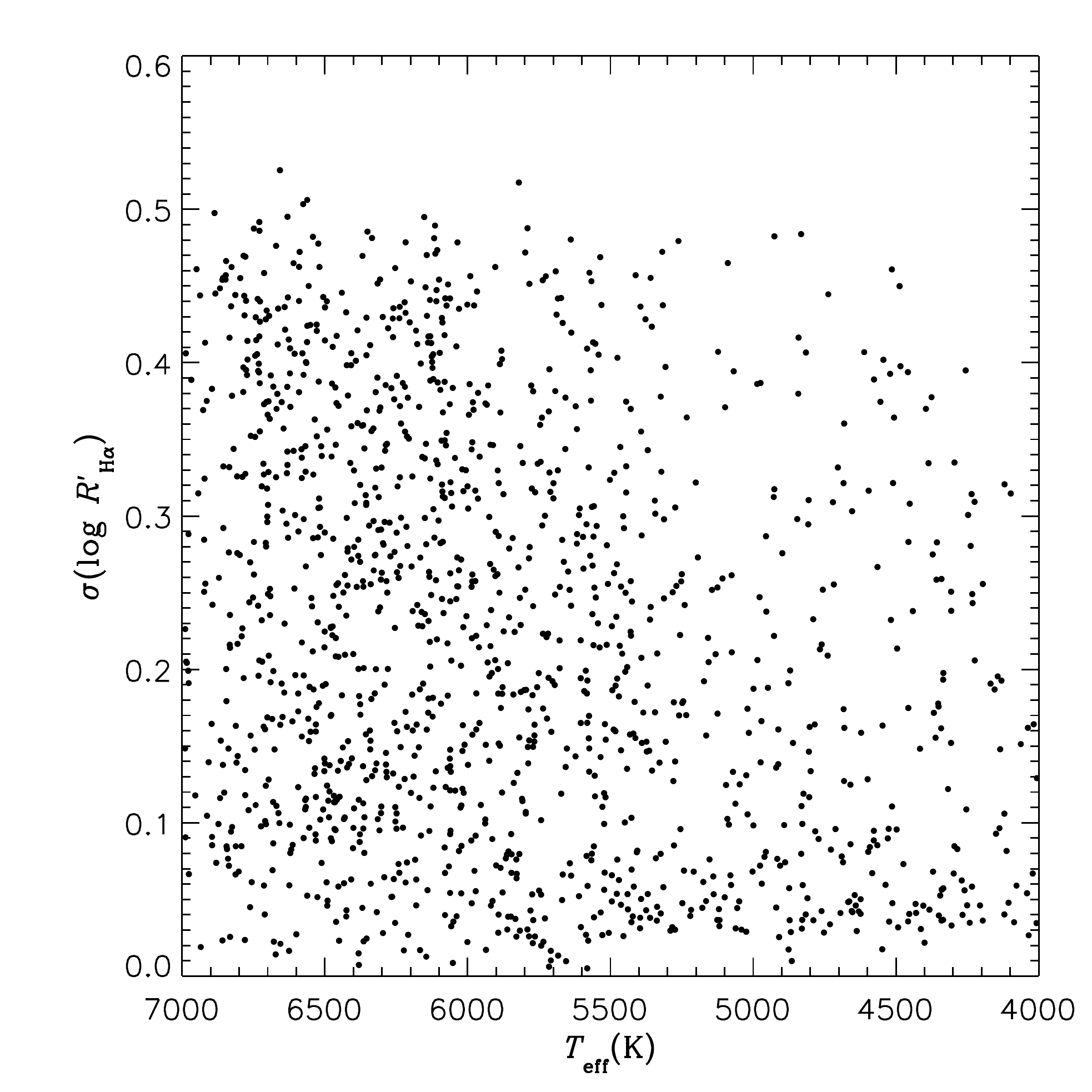}{0.5\textwidth}{(d)}
          }
\caption{(a): $\rm \sigma(EW_{CaK})$ vs. $T\rm_{eff}$. (b): $\rm \sigma(EW_{H\alpha})$ vs. $T\rm_{eff}$. (c): $\rm \sigma(\log R'_{CaK})$ vs. $T\rm_{eff}$. (d): $\rm \sigma(\log R'_{H\alpha})$ vs. $T\rm_{eff}$.    \label{fig: EWerr_Teff}}
\end{figure}

Monte Carlo simulation is used to obtain the error of $\rm EW$. For a spectrum, flux at every data point has a inverse variance. So the random flux can be produced following a gaussian distribution of $\mu$(flux at a data point) and $\sigma$(inverse variance). In our simulation, we produce 1,000 simulated spectra and calculate their $\rm EW$. The standard deviation of $\rm EW$ is used as the error of $\rm EW$. Figure \ref{fig: EWerr_Teff}(a) and (b) plots $\rm \sigma(EW)$ vs. $T\rm_{eff}$. From Figure \ref{fig: EWerr_Teff}(a), the error of $\rm EW_{CaK}$ increases slightly as $T\rm_{eff}$ decreases. The average error of $\rm EW_{CaK}$ is about 0.1\AA\ when $T\rm_{eff}>5500$K and about 0.2\AA\ at $T\rm_{eff}=4500$K. From Figure \ref{fig: EWerr_Teff}(b), we see that the bottom boundary of distribution of $\rm \sigma(EW_{H\alpha})$ increases slightly when $T\rm_{eff}$ decreases from 5500K. The average error of $\rm EW_{H\alpha}$ is about 0.04\AA.

The error of $\log R'$ is also estimated by using Monte Carlo simulation again. Only four factors are considerd: the $\rm EW$ and stellar atmospheric parameters including $T\rm_{eff}$, [Fe/H] and $\log g$. Errors of $T\rm_{eff}$, [Fe/H] and $\log g$ are set to 110K, 0.12 dex and 0.11 dex, respectively. Figure \ref{fig: EWerr_Teff}(c) and (d) shows $\rm \sigma(\log R')$ vs. $T\rm_{eff}$. From Figure \ref{fig: EWerr_Teff}(c), we see that when $T\rm_{eff}>6000$K, $\rm \sigma(\log R'_{CaK})$ shows a large scatter. The large scatter is mainly due to small $\rm EW'_{CaK}$ which means $\rm EW_{CaK}$ are close to the basal line. If a star has $\rm EW_{CaK}$ close to the basal line, its $\rm EW'_{CaK}$ is close to zero. A small difference in $\rm EW_{CaK}$ can cause a large difference in $\rm \log R'_{CaK}$ when taking logarithm of $R'$ ($R'\rm=EW'\times \chi$). When $T\rm_{eff}<6000$K, $\rm \sigma(\log R'_{CaK})$ shows a relatively tight distribution. The errors are about 0.05dex and 0.15dex at 5500K and 4500K. There is no large scatter at this $T\rm_{eff}$ range because as $T\rm_{eff}$ decreases the distribution of $\rm EW_{CaK}$ shows a large scatter (see Figure \ref{fig: EW_Teff}) and many stars have relatively large $\rm EW'_{CaK}$. From Figure \ref{fig: EWerr_Teff}(d), we see that at all $T\rm_{eff}$ range $\rm \sigma(\log R'_{H\alpha})$ shows a very large scatter from 0.0dex to 0.5dex. The reason is same as above. However, for the H$\alpha$, many member stars have $\rm EW_{H\alpha}$ close to the basal line not only at high $T\rm_{eff}$ range but also at low $T\rm_{eff}$ range.

\section{the impact of binaries and interstellar medium on $\log R'$}\label{binaries and interstellar medium}

The interaction of two stars can affect CA level. The member stars list provided by \cite{Gaudin2018} include gaia color ($G_{BP}-G_{RP}$) and visual magnitude ($G_{mag}$). According to these information, we can plot CMD (color magnitude diagram) of each open cluster. On CMD, some member stars lie above the single main sequence and many of them are binaries. We try to check change of mean value and scatter of $\log R'$ by discarding these stars. For example, after discarding these stars of NGC 2632, the mean value and standard error of $\log R\rm'_{CaK}$ changes from -5.04$\pm$0.184 to -5.05$\pm$0.184 with 4000K $<T\rm_{eff}<$ 5500K. The same value of $\log R\rm'_{H\alpha}$ changes from -5.16$\pm$0.185 to -5.18$\pm$0.178 with 4000K $<T\rm_{eff}<$ 5500K. Note that we do not consider the labels provided by simbad. The change of the mean values is very small.

\begin{figure}
\center
\includegraphics[scale=0.7]{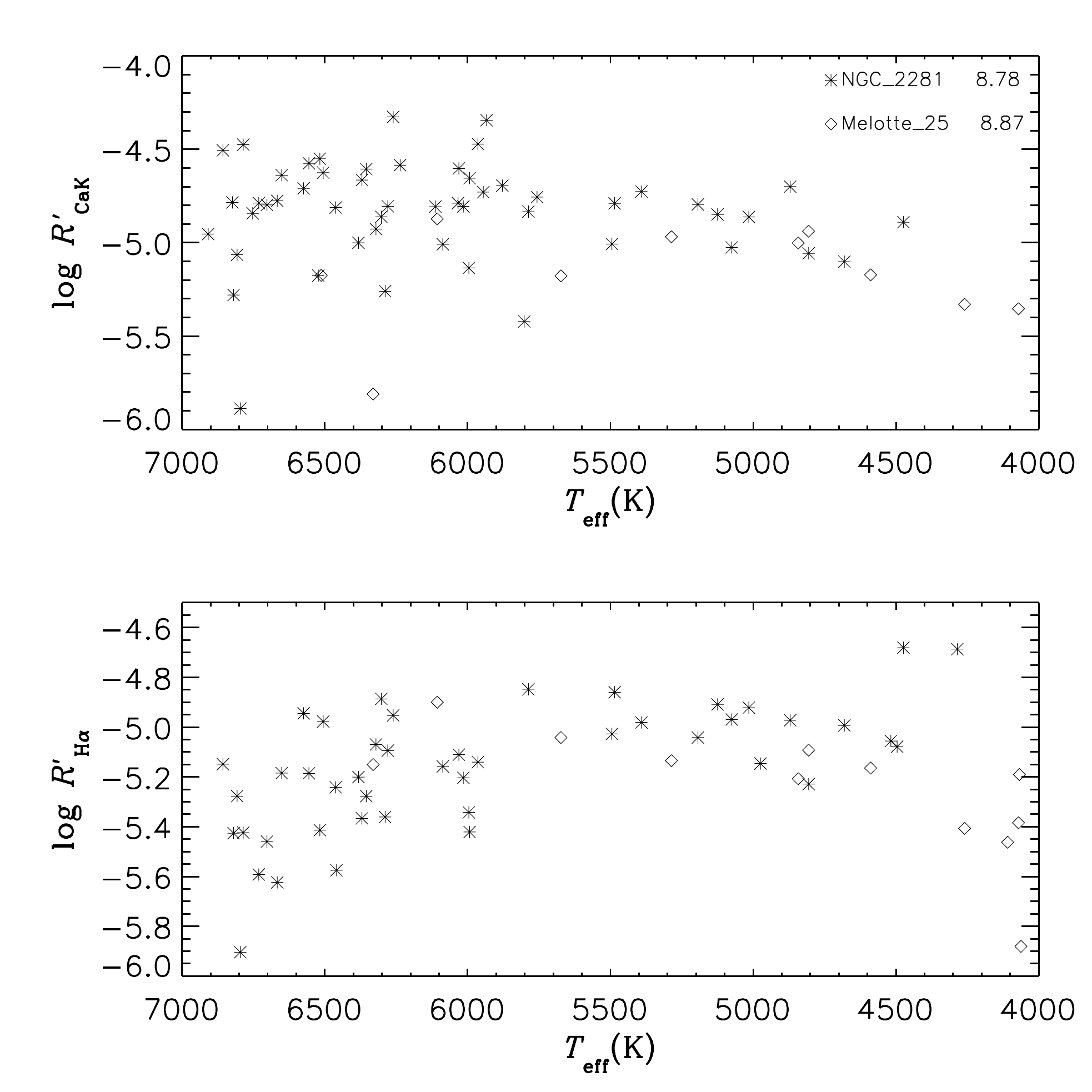}
\caption{Top panel is $\log R\rm'_{CaK}$ vs. $T\rm_{eff}$. Bottom panel is $\log R'\rm_{H\alpha}$ vs. $T\rm_{eff}$. The asterisk represents for NGC 2281 while the diamond represents for Melotte 25. \label{fig:ISM}}
\end{figure}

The interstellar medium (ISM) imprints absorption lines in the vicinity of the CaII H \& K line cores, which negatively biases CA indice \citep{PACE2004,Curtis2017}. Our spectra are low resolution spectra (R $\sim$ 1800 at 5500\AA) and they are not much likely to show evident ISM absorption lines at the wavelength of the CaII H \& K line. H$\alpha$ line is less affected by ISM. So we can find some open clusters which are coeval but separated by a large distance. Then we plot the distributions of $\log R\rm'_{CaK}$ vs. $T\rm_{eff}$ and $\log R\rm'_{H\alpha}$ vs. $T\rm_{eff}$. If ISM affect our results, at a similar range of $T\rm_{eff}$, the $\log R\rm'_{CaK}$ values of nearby cluster suppose to be higher than that of distant cluster, but the $\log R\rm'_{H\alpha}$ values should keep consistent. However, in our sample, the number of member stars for some open clusters are small. Besides, many open clusters have most member stars with $T\rm_{eff}>6000$K. When $T\rm_{eff}>6000$K, $\log R'$ values of member stars of different ages can mix. Fortunately, we find two open clusters: NGC 2281 and Melotte 25 (Hyades). NGC 2281 has $\log t = 8.78$ \citep{Kha2013} and $d = 519$pc \citep{Gaudin2018}. $d$ is the distance to the sun. Melotte 25 has $\log t = 8.87$ \citep{Gos2018} and $d = 48$pc \citep{Roser2019}. There is more than 100Myr age difference between the two clusters. The distributions of $\log R'$ vs. $T\rm_{eff}$ are shown in Figure \ref{fig:ISM}. When $T\rm_{eff}<6000$K, NGC 2281 has a little larger $\log R\rm'_{CaK}$ and $\log R\rm'_{H\alpha}$ values than that of Melotte 25 at a similar range of $T\rm_{eff}$, which suggests that the impact of ISM is smaller compared to the decrease in CA over time.

\section{An illustration of logarithm effect}\label{An illustration of logarithm effect}

During data processing, $\rm EW$ of some member stars are close to the basal lines. For those stars, a small difference in $\rm EW$ can cause a large difference in $\log R'$. This is because $\rm EW$ close to the basal line leads to $\rm EW'$ close to zero and $R'$ close to zero ($R'\rm=EW'\times \chi$), then $R'$ is projected to a large range when taking logarithm. We list some examples in Table \ref{Table: Examples to show logarithm effect} to illustrate it.

\startlongtable
\begin{deluxetable}{ccccc}
\tablecaption{Some examples to illustrate logarithm effect\label{Table: Examples to show logarithm effect}}
\tablehead{
\colhead{$T\rm_{eff}$(K)} & \colhead{$\rm EW_{CaK}$(\AA)}  & \colhead{$\log R'_{\rm CaK}$}  & \colhead{$\rm EW_{H\alpha}$(\AA)}
& \colhead{$\log R'_{\rm H\alpha}$}
}
\startdata
  6500 & -4.50 & -4.92 & -2.50 & -5.75\\
  6500 & -4.00 & -4.46 & -2.00 & -4.77\\
  6500 & -3.50 & -4.24 & -1.50 & -4.49\\
  5500 & -5.00 & -5.42 & -1.70 & -5.74\\
  5500 & -4.50 & -4.83 & -1.20 & -4.72\\
  5500 & -4.00 & -4.59 & -0.70 & -4.44\\
  4500 & -4.50 & -5.59 & -0.70 & -5.40\\
  4500 & -4.00 & -5.25 & -0.20 & -4.71\\
  4500 & -3.50 & -5.06 & 0.30 & -4.46\\
\enddata
\tablecomments{Note that [Fe/H] and $\log g$ are set to 0.0 and 4.2.}
\end{deluxetable}

\end{document}